\documentclass[useAMS,usenatbib,usegraphicx]{mn2e}
\usepackage{amssymb,amsmath, color, soul, ulem}

\newcommand{\msano}{{\rm M}_\odot ~{\rm yr}^{-1}}
\newcommand{\alf}{Alfv\'en}
\newcommand{\jdot}{\dot{J}}
\newcommand{\mdot}{\dot{M}}

\title[M-dwarf stellar winds]{M-dwarf stellar winds: the effects of realistic magnetic geometry on rotational evolution and planets}
\author[A.~A.~Vidotto et al.]{A.~A.~Vidotto$^{1}$\thanks{E-mail: Aline.Vidotto@st-andrews.ac.uk}, {M.~Jardine}$^{1}$,  {J.~Morin}$^{2,3,4}$,  {J.~F.~Donati}$^{5}$, {M.~Opher}$^{6}$, {T.~I.~Gombosi}$^{7}$ \\
$^{1}$SUPA, School of Physics and Astronomy, University of St Andrews, North Haugh, St Andrews, KY16 9SS, UK\\
$^{2}$Institut f\"ur Astrophysik, Georg-August-Universit\"at, Friedrich-Hund-Platz 1, D-37077, Goettingen, Germany\\
$^{3}$LUPM-UMR5299, Universit\'e Montpellier II \& CNRS, Place Eug\`ene Bataillon,
F-34095 Montpellier Cedex 05, France \\
$^{4}$Dublin Institute for Advanced Studies, School of Cosmic Physics, 31 Fitzwilliam Place, Dublin 2, Ireland \\
$^{5}$LATT - CNRS/Universit\'e de Toulouse, 14 Av.~E.~Belin, Toulouse, F-31400, France\\
$^6$Boston University, 725 Commonwealth Ave, Boston, MA, 02215, USA\\
$^{7}$University of Michigan, 1517 Space Research Building, Ann Arbor, MI, 48109-2143, USA}

\begin{document}

\date{Accepted . Received ; in original form}

\pagerange{\pageref{firstpage}--\pageref{lastpage}} \pubyear{2013}

\maketitle

\label{firstpage}

\begin{abstract}
We perform three-dimensional numerical simulations of stellar winds of early-M dwarf stars. Our simulations incorporate observationally reconstructed large-scale surface magnetic maps, suggesting that the complexity of the magnetic field can play an important role in the angular momentum evolution of the star, possibly explaining the large distribution of periods in field dM stars, as reported in recent works. In spite of the diversity of the magnetic field topologies among the stars in our sample, we find that stellar wind flowing near the (rotational) equatorial plane carries most of the stellar angular momentum, but there is no preferred colatitude  contributing to mass loss, as the mass flux is maximum at different colatitudes for different stars. We find that more non-axisymmetric magnetic fields result in more asymmetric mass fluxes and wind total pressures $p_{\rm tot}$ (defined as the sum of thermal, magnetic and ram pressures). Because planetary magnetospheric sizes are set by pressure equilibrium between the planet's magnetic field and $p_{\rm tot}$, variations of up to a factor of $3$ in $p_{\rm tot}$ (as found in the case of a planet orbiting at several stellar radii away from the star) lead to variations in magnetospheric radii of about 20 percent along the planetary orbital path. In analogy to the flux of cosmic rays that impact the Earth, which is inversely modulated with the non-axisymmetric component of the total open solar magnetic flux, we conclude that planets orbiting M dwarf stars like DT~Vir, DS~Leo and GJ~182, which have significant non-axisymmetric field components, should be the more efficiently shielded from galactic cosmic rays, even if the planets lack a protective thick atmosphere/large magnetosphere of their own. 
\end{abstract}

\section{INTRODUCTION}\label{sec.intro}
Stellar winds are believed to explain the observed rotational braking of main-sequence stars with outer convective envelopes (spectral types later than mid-F), as they act as an efficient removal mechanism for the star's angular momentum \citep[e.g.,][]{1962AnAp...25...18S, 1967ApJ...148..217W,1968MNRAS.138..359M,1976ApJ...210..498B}. Because mass loss takes place throughout the star's main-sequence life, as the stars age, they spin down. For solar-mass main-sequence stars, it has been both empirically recognised \citep{1972ApJ...171..565S} and theoretically shown \citep[e.g.,][]{1987MNRAS.226...57M} that the stellar angular rotation velocity $\Omega_*$ scales with the age $t$ of the star as $\Omega_\star \propto t^{-1/2}$. 

By comparing colour-period diagrams for open clusters at different ages, it has been suggested that, as the cluster ages, stars that were once part of the spread group of fast rotators evolve into the more defined sequence of slow/moderate rotators as they spin-down \citep{2003ApJ...586..464B, 2010ApJ...721..675B}. This evolution appears to occur more rapidly for more massive stars, indicating that the time for a star to spin-down increases with decreasing stellar mass. \citet{2011ApJ...733..115M} estimate that G dwarfs should make this transition in a time scale $\lesssim 150$~Myr, while early to mid-K dwarfs should take $150 - 300$~Myr, and late K dwarfs would take $\sim 300 - 600$~Myr to evolve from the sequence of fast rotators to the moderate/slow rotator sequence. 

If the same trend continues down to the lower-mass-object range, one expects the spin-down time scale to be even longer for M dwarf (dM) stars. Indeed, \citet{2011MNRAS.413.2218D} showed that by the Hyades age ($\sim 625$~Myr), dM stars with masses above $0.5~M_\odot$ have already converged towards the tight period--colour sequence (also shown as period--mass relation). By investigating a sample of field-dM stars in the fully convective regime (masses $\lesssim 0.35~M_\odot$), \citet{2011ApJ...727...56I} showed that kinematically young (thin disc, ages $\sim $~Gyr) objects rotate faster than the kinematically old (thick disc, ages $\sim 7 - 12$~Gyr) objects, in agreement with previous expectations and also with activity lifetimes of late-dM stars \citep[$\approx 7$~Gyr]{2008AJ....135..785W}. 


None the less, recent studies have revealed an interesting behaviour for field dM stars less massive than $0.55 ~M_\odot$ (\citealt{2011ApJ...727...56I} from Mearth data; \citealt{2013MNRAS.432.1203M} from Kepler data). First, a wide range of rotation rates is observed, where a population of rapidly rotating dM stars coexist with a significant population of slow rotators (possibly due to populations of stars with different ages). Second, a trend exists in the upper envelope of the period--mass relation, which changes sign at masses around $0.55 ~M_\odot$. Note that the upper envelope of the perio--mass relation is defined by the slowest rotating stars at each mass bin. For $M_\star \lesssim 0.55 ~M_\odot$, \citet{2013MNRAS.432.1203M} found that the period of the slowest rotating objects rises with decreasing mass. This indicates that some dM stars might lose angular momentum more efficiently than other dM stars and higher mass stars. These observations represent a challenge for models of rotational evolution cool stars. New analytical models have been derived by \citet{2012ApJ...746...43R}, assuming a monopolar magnetic field, whose intensity does not depend on the stellar mass nor time. Within this framework, the increase of the braking time scale toward decreasing mass observed close to the fully convective boundary naturally arises from the strong decrease of stellar radii in this mass range. In addition, they argue that the existence of slowly rotating fully-convective field M dwarfs, could be accounted for by assuming that the rotation rate $\Omega_{\rm sat}$ at which activity saturation occurs on lower-mass stars is much larger than $\Omega_{\rm sat}$  of higher mass stars. We will argue in Section~\ref{sec.rotev} that taking into account the topology of the magnetic field might be an alternative (and perhaps additional) explanation.

In order to reproduce rotational evolution of stars in open clusters,  empirically-motivated revisions in the standard solar wind-prescription have been used \citep{1997A&A...326.1023B,2009IAUS..258..363I,2012ApJ...746...43R,2013A&A...556A..36G}. These modifications, for example, allow for the presence of different magnetic field geometries \citep{1988ApJ...333..236K}, angular velocity saturation \citep{1987ApJ...318..337S, 1996ApJ...462..746B}, decoupling between the radiative core and the convective envelope \citep{1991ApJ...376..204M}. The downside of the inclusion of empirical phenomena in such prescriptions is that they introduce parameters that are arbitrarily adjusted to fit the data.  It is, therefore, crucial to understand from theoretical principles the role that winds of low-mass stars play on the extraction of angular momentum. Recent works have provided important steps towards that direction \citep{2012ApJ...746...43R,2012ApJ...754L..26M}, but the role of different magnetic topologies has not yet been investigated.  

The magnetic field that emerges at the surface of the stars is expected to present different characteristics, reflecting the different stellar internal structures and operating dynamo mechanisms. This expectation is confirmed by recent surveys that probe the large-scale topology of the surface magnetic fields of dM stars \citep{2008MNRAS.390..567M,2010MNRAS.407.2269M,2008MNRAS.390..545D}. \citet{2008MNRAS.390..567M} showed that mid-dM stars (either fully convective or with a small radiative core) exhibit strong poloidal axisymmetric dipole-like surface magnetic topologies, while the partially convective ones present weaker, non-axisymmetric fields with significant toroidal component \citep{2008MNRAS.390..545D}. The picture that arises from the analysis of a more recent sample of late-M objects \citep{2010MNRAS.407.2269M} shows two distinct populations: one with very strong axisymmetric poloidal fields (similar to the mid-M stars) and another with significant non-axisymmetric component, plus a significant toroidal component.

The effect that complex magnetic field topologies might have on stellar winds has not been systematically investigated in the literature, as most of the theoretical work developed so far relies on simplified, axisymmetric geometries for the magnetic field. \citet{2010ApJ...720.1262V} performed simulations of stellar winds of young Suns with different alignments between the rotation axis and the magnetic dipole axis. In those simulations, angular momentum losses are enhanced by a factor of $2$ as one goes from the aligned case to the case where the dipole is tilted by $90^{\circ}$ (i.e., as non-axisymmetry is increased).  In the present work, the variations of surface characteristics are observationally determined and constitute therefore an extra step towards more realistic models of stellar winds of low-mass stars. 

To investigate the behaviour of angular momentum loss of low-mass stars, we present in this paper a comparative study of stellar winds of a sample of early-dM stars  (spectral types M0 to M2.5), for which surface magnetic field maps have been obtained. Section~\ref{sec.sample} presents our sample of stars. Sections~\ref{sec.model} and \ref{sec.results} describe the numerical model used in the simulations and their results, respectively. A discussion about the effects of the field topology on angular momentum losses are presented in Section~\ref{sec.discussion}, where we also discuss possible effects on orbiting planets. In Section~\ref{sec.conclusions}, we present our summary and conclusions of this work.

\section{Sample of stars}\label{sec.sample}
\begin{figure*}
\includegraphics[width=56mm]{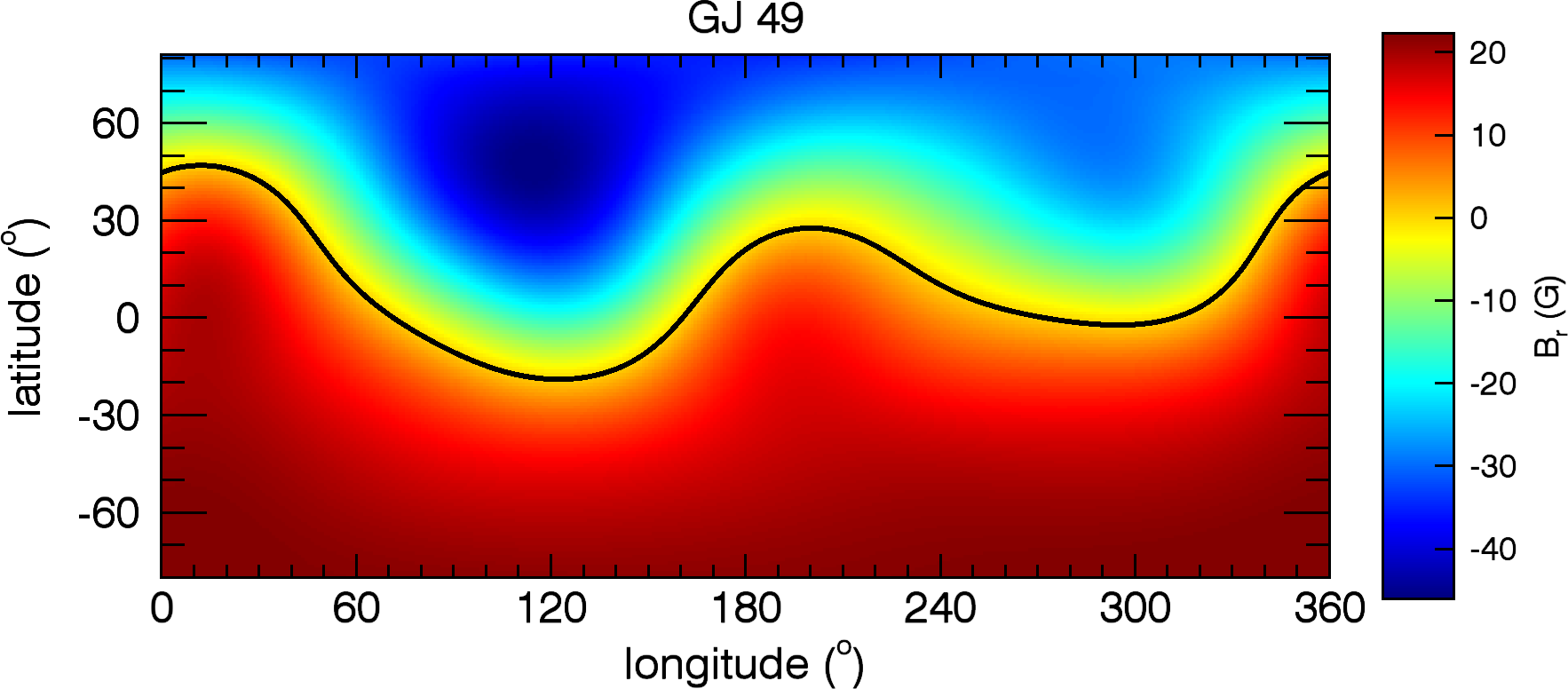}
\includegraphics[width=56mm]{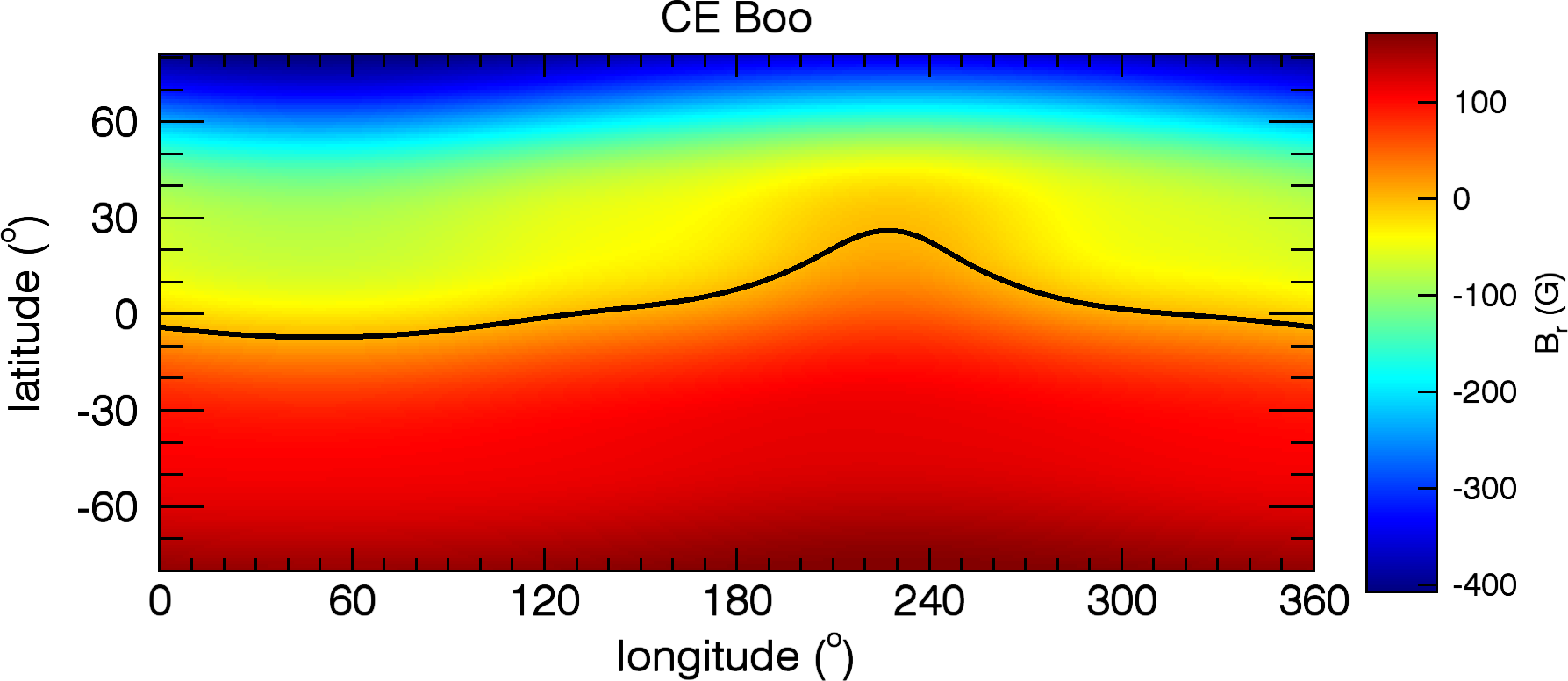}
\includegraphics[width=56mm]{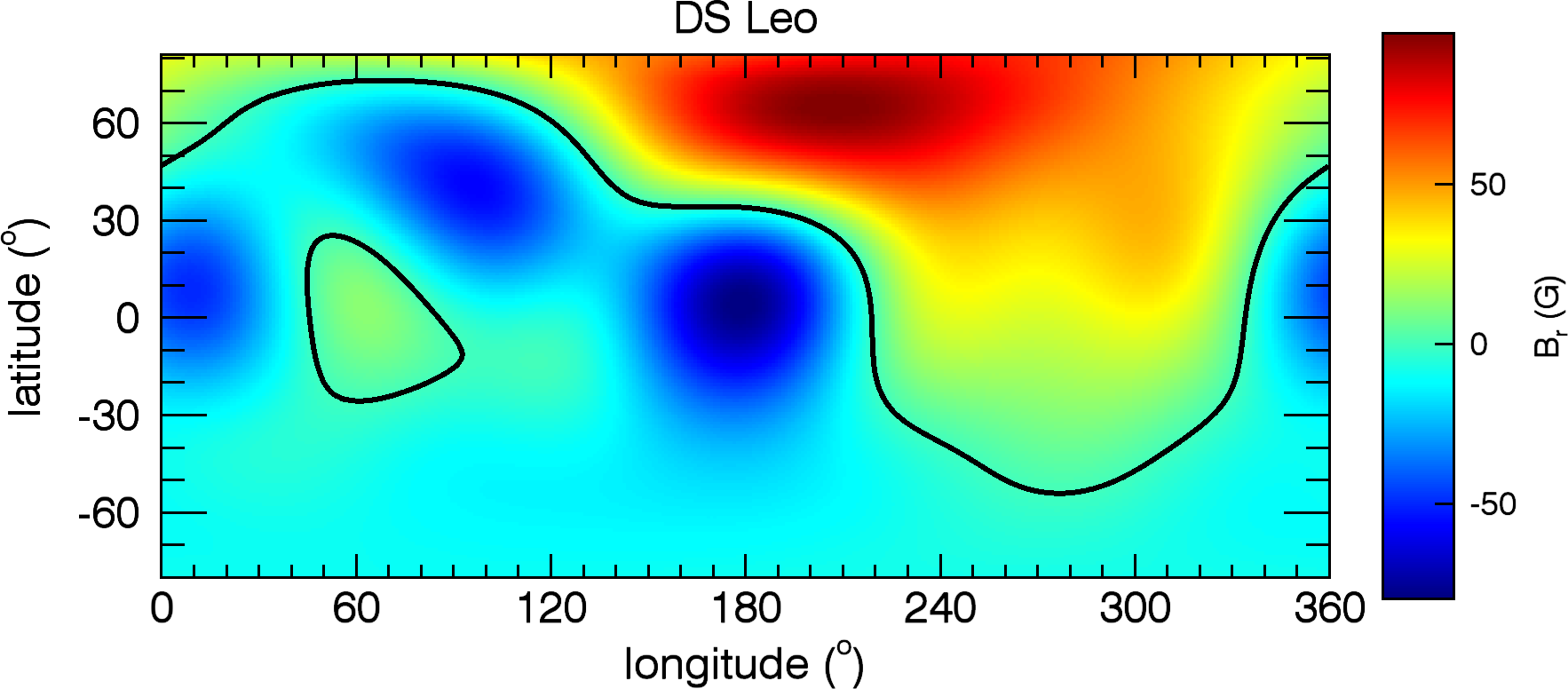}\\
\includegraphics[width=56mm]{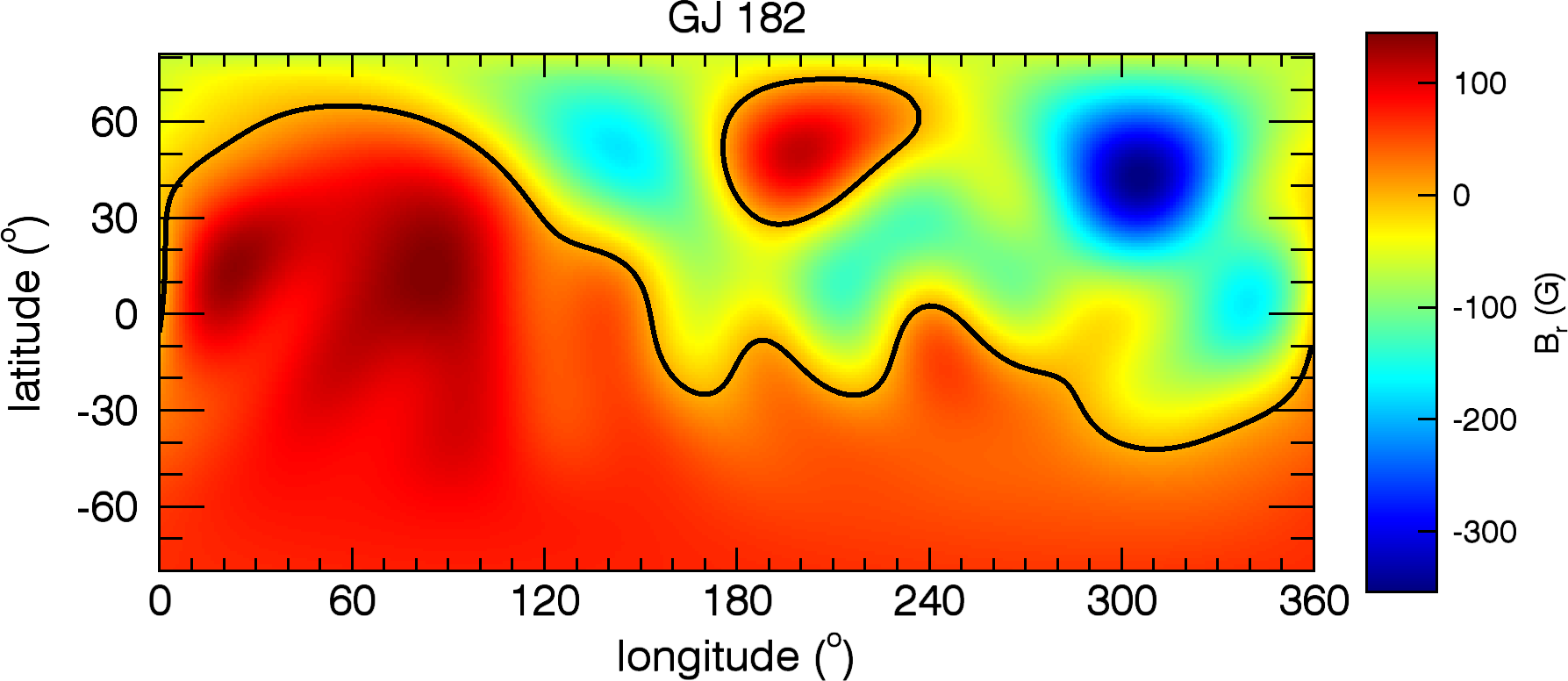}
\includegraphics[width=56mm]{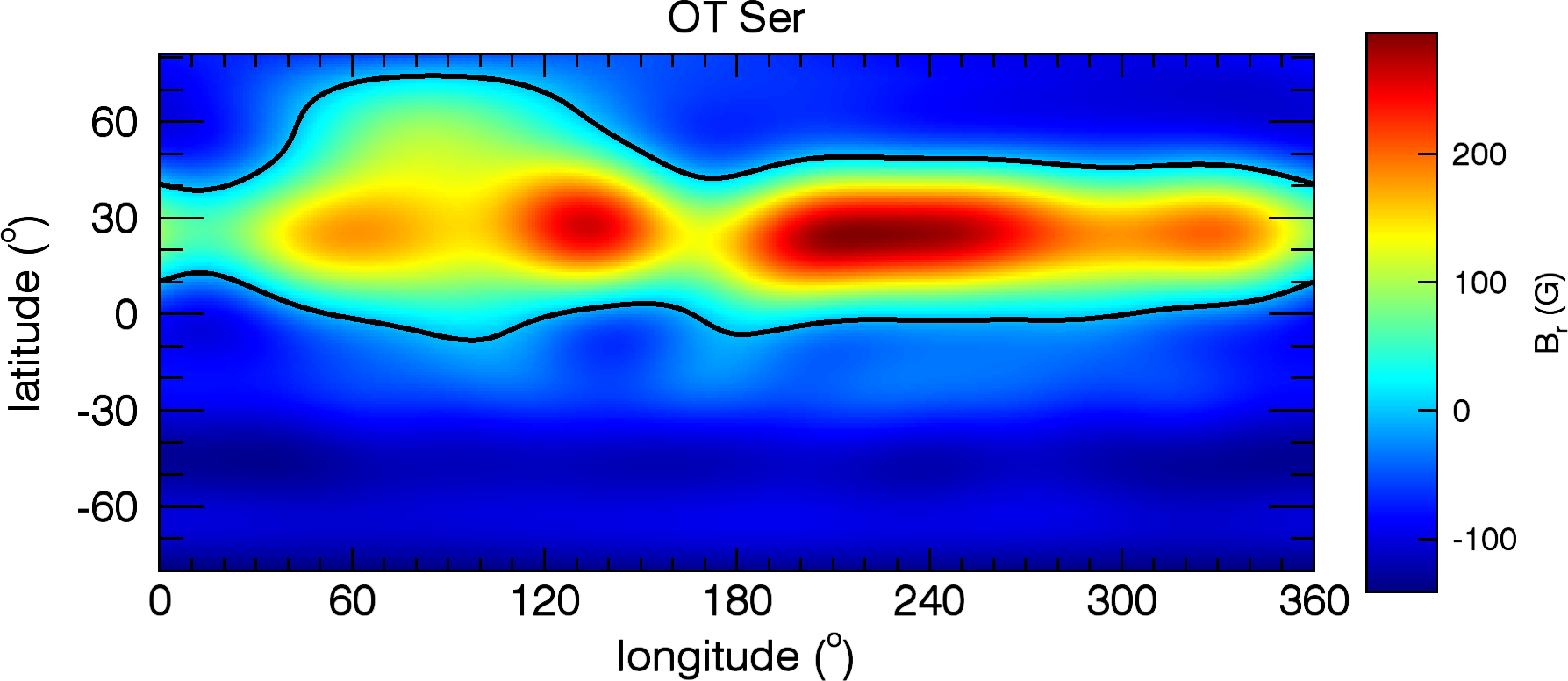}
\includegraphics[width=56mm]{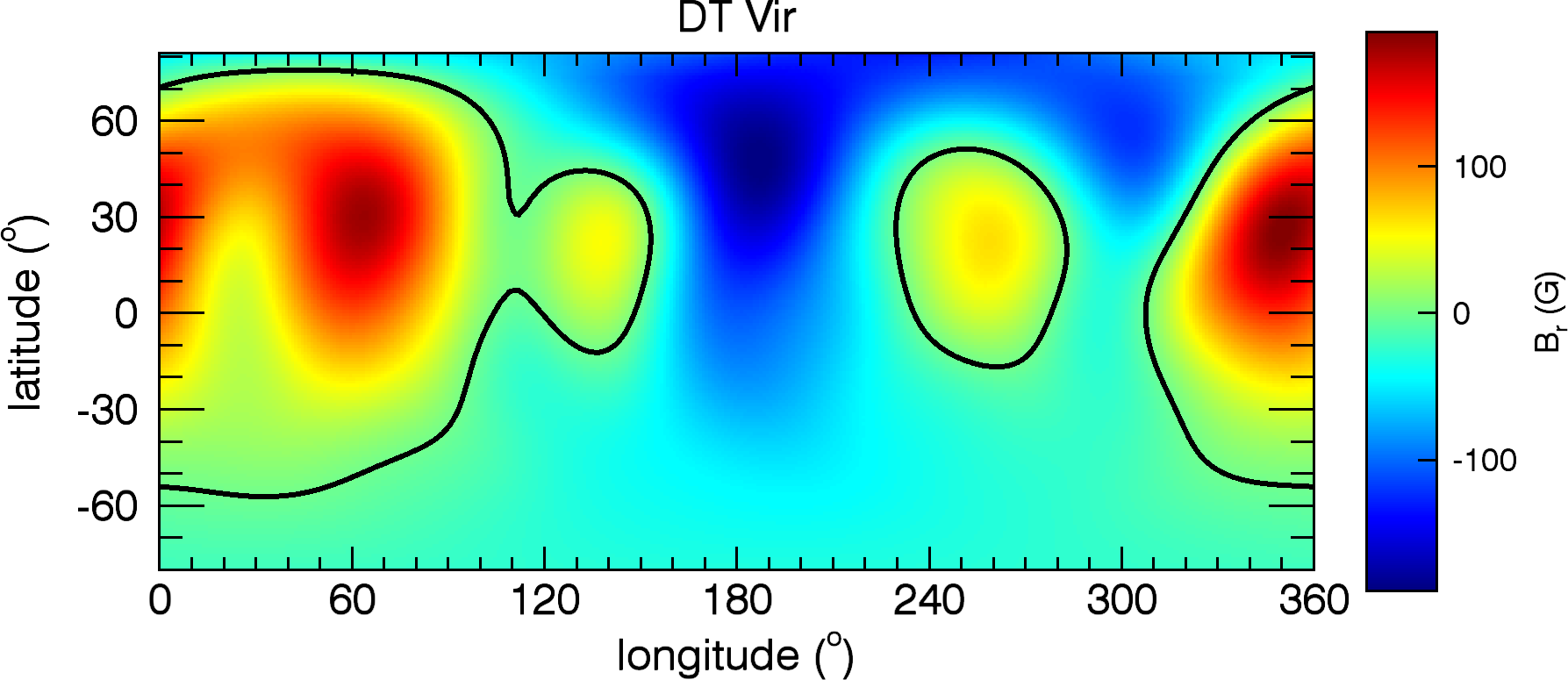}\\
\caption{The radial component of the observationally reconstructed magnetic field of the stars in our sample \citep{2008MNRAS.390..545D}. The black solid line shows an iso-contour of $B_r=0$. \label{fig.maps}}
\end{figure*}

The stars considered in this study consist of six early-M stars (spectral types M0 to M2.5), for which the large-scale surface magnetic field maps have been reconstructed from a series of circular polarisation spectra using the Zeeman-Doppler Imaging (ZDI) technique \citep[e.g.,][]{1997A&A...326.1135D, 2012EAS....57..165M}. In this work, we concentrate on the early-dM stars and use the maps that were published in \citet{2008MNRAS.390..545D}. Figure~\ref{fig.maps} presents the reconstructed surface field of these stars and Table \ref{table} presents a summary of their characteristics. Our targets, namely GJ 49, DS Leo, DT Vir, OT Ser, GJ 182 and CE Boo, present more complex surface magnetic fields than V374 Peg \citep{2006Sci...311..633D}, a mid-dM star that was investigated in a previous model \citep{2011MNRAS.412..351V}.  

\begin{table*} 
\centering
\caption{Characteristics of our sample of stars. The columns are, respectively: the star name, the observation epoch, the stellar spectral type,  mass $M_\star$, radius $R_\star$, rotation period $P_{\rm rot}$, the unsigned surface magnetic fluxes ($\Phi_0$, Eq.~(\ref{eq.phi0})), and the fractional energy in the poloidal axisymmetric modes ($f_{\rm axi}$). Values are from \citet{2008MNRAS.390..545D}. \label{table}}    
\begin{tabular}{llccccccccccccccccc}  
\hline
Star	&	Obs.	&	Sp.		&	$	M_\star 	$	&	$	R_\star		$	&	$P_{\rm rot}	$	&	$	\Phi_0		$	& $f_{\rm axi}$ \\ 	
ID		&	Epoch	&	Type &	$(M_\odot)	$	&	$(R_\odot)	$		&	$	(d)	$		&$	(10^{23}~{\rm Mx})	$& \\ 
 \hline	
      GJ 49 &Jul/07 &  M1.5  & $0.57$ & $0.51$ & $18.6$ & $  2.6$ & $ 0.67 $ \\
    CE Boo &Jan/08 &  M2.5  & $0.48$ & $0.43$ & $14.7$ & $   11$ & $ 0.96 $ \\
    DS Leo &Dec/07 &    M0  & $0.58$ & $0.52$ & $14.0$ & $  3.9$ & $ 0.16 $ \\ \hline
    GJ 182 &Jan/07 &  M0.5  & $0.75$ & $0.82$ & $4.35$ & $   30$ & $ 0.17 $ \\ 
    OT Ser &Jul/07 &  M1.5  & $0.55$ & $0.49$ & $3.40$ & $   13$ & $ 0.86 $ \\
    DT Vir &Jan/07 &  M0.5  & $0.59$ & $0.53$ & $2.85$ & $  9.4$ & $ 0.12 $ \\
 \hline			
\end{tabular}
\end{table*}

 We note that these six stars comprise two groups with similar characteristics in the mass--period diagram, as can be seen in Figure~14 of \citet{2008MNRAS.390..545D}. In the first group, GJ 49, DS Leo and CE Boo have rotational periods of $\sim 15$~days, while, in the second group, DT Vir, OT Ser, and GJ 182 rotate faster with periods of $\sim 3$~days. They all present similar masses. Despite sharing similar characteristics, members of each group present different surface magnetic field topologies and intensities. For this reason, this sample is useful for investigating the effects that different magnetic field characteristics play on stellar winds. To account for the observed three-dimensional (3D) nature of their magnetic fields, 3D stellar wind models are required.

\section{Stellar wind model}\label{sec.model}
To simulate the stellar winds of the dM stars in our sample, we use the 3D magnetohydrodynamics (MHD) numerical code BATS-R-US \citep{1999JCoPh.154..284P}. BATS-R-US solves the ideal MHD equations
\begin{equation}
\label{eq:continuity_conserve}
\frac{\partial \rho}{\partial t} + \boldsymbol\nabla\cdot \left(\rho {\bf u}\right) = 0,
\end{equation}
\begin{equation}
\label{eq:momentum_conserve}
\frac{\partial \left(\rho {\bf u}\right)}{\partial t} + \boldsymbol\nabla\cdot\left[ \rho{\bf u\,u}+ \left(P + \frac{B^2}{8\pi}\right)I - \frac{{\bf B\,B}}{4\pi}\right] = \rho {\bf g},
\end{equation}
\begin{equation}
\label{eq:bfield_conserve}
\frac{\partial {\bf B}}{\partial t} + \boldsymbol\nabla\cdot\left({\bf u\,B} - {\bf B\,u}\right) = 0,
\end{equation}
\begin{equation}
\label{eq:energy_conserve}
\frac{\partial\varepsilon}{\partial t} +  \boldsymbol\nabla \cdot \left[ {\bf u} \left( \varepsilon + P + \frac{B^2}{8\pi} \right) - \frac{\left({\bf u}\cdot{\bf B}\right) {\bf B}}{4\pi}\right] = \rho {\bf g}\cdot {\bf u} ,
\end{equation}
where the eight primary variables are the mass density $\rho$, the plasma velocity ${\bf u}=\{ u_r, u_\theta, u_\varphi\}$, the magnetic field ${\bf B}=\{ B_r, B_\theta, B_\varphi\}$, and the gas pressure $P$. The gravitational acceleration due to the star with mass $M_\star$ and radius $R_\star$ is given by ${\bf g}$, and $\varepsilon$ is the total energy density given by 
\begin{equation}\label{eq:energy_density}
\varepsilon=\frac{\rho u^2}{2}+\frac{P}{\gamma-1}+\frac{B^2}{8\pi} ,
\end{equation}
where $\gamma$ is the polytropic index ($p\propto \rho^\gamma$). We consider an ideal gas, so $P=n k_B T$, where  $k_B$ is the Boltzmann constant, $T$ is the temperature, $n=\rho/(\mu m_p)$ is the particle number density of the stellar wind, $\mu m_p$ is the mean mass of the particle. In this work, we adopt $\gamma=1.1$ and $\mu=0.5$. The physical processes that are responsible for heating the solar corona and accelerating the solar wind are not yet known. \citet{2009LRSP....6....3C} provides a recent review on the solar wind acceleration, which has been attributed to waves and turbulence in open flux tubes \citep[e.g.,][]{1973ApJ...181..547H,2006JGRA..111.6101S,2007ApJS..171..520C}, magnetic reconnection events \citep[e.g.,][]{1999JGR...10419765F,2006ApJ...642.1173S}, etc \citep[see also][]{2007RvGeo..45.1004M}. In spite of the current lack of a complete theoretical model (i.e., from first principle Physics starting at a photospheric level upwards to the corona) for the acceleration of the solar wind, empirical correlations have been used to  predict the solar wind characteristics at different distances \citep{1990ApJ...365..372W, 2000JGR...10510465A, 2007ApJ...654L.163C, 2012ApJ...756..155E}. These correlations are constrained by direct observation of the solar wind. It is however not clear how they could be applied to winds of different solar-type stars, since the lack of direct observations of solar-like winds prevent a direct scaling of the empirical correlations observed in the solar wind. Although having similar masses, radii and effective temperatures, solar-like stars might have considerable different characteristics than those of the Sun and that can affect their wind properties. The variety of observed rotation rates, intensities and topologies of their magnetic fields, X-ray luminosities and coronal temperatures of cool, dwarf stars suggest that their winds might be different from the solar one \citep[see discussion in][]{2013arXiv1310.0593V}. In the present work, we adopt a simplified wind-driving mechanism, where we assume the wind to be a polytrope, where the polytropic index $\gamma$ is a free parameter of the model. We keep the same $\gamma$ for all the simulations, so as to have a homogeneous parameter space for all the cases studied. The wind solutions we found could to be affected if a different acceleration mechanism is chosen, but it is beyond the scope of this paper to perform such an investigation.

At the initial state of the simulations, we assume that the wind is thermally driven \citep{1958ApJ...128..664P}. The stellar rotation period $P_{\rm rot}$, $M_\star$ and $R_\star$ are given in Table \ref{table}. At the base of the corona ($r=R_\star$), we adopt a wind coronal temperature $T_0 = 2\times 10^6$~K and wind number density $n_0=10^{11}$cm$^{-3}$. With this numerical setting, the initial solution for the density, pressure (or temperature) and wind velocity profiles are fully specified. Note that the wind base density is an unconstrained input parameter of global stellar wind models. To better constrain the coronal base density, more precise measurements of mass-loss rates of dM stars are desired. However, mass-loss rates of cool stars are notoriously difficult to observe \citep[e.g.,][]{2005ApJ...628L.143W}. Traditional mass-loss signatures, such as P Cygni line profiles, are not  observed in dM stars due to the optically-thin nature of their stellar winds. Estimates of mass-loss rates of dM stars in the literature are rather controversial and span more than five orders of magnitude, ranging from a subsolar value of $\dot{M} \simeq 4 \times 10^{-15}~\msano$ \citep{2001ApJ...547L..49W} to supersolar values of $\dot{M} \simeq 10^{-10}~\msano$  \citep{1992ApJ...397..225M}. The values of $\dot{M}$ derived from our simulations fall in the range of $\mdot$ predicted by several estimates, but it is still not properly constrained. 

To complete our initial numerical set up, we incorporate the radial component of the magnetic field $B_r$ reconstructed from observations using the ZDI technique (Fig.~\ref{fig.maps}). This is similar to the method presented in \citet{2012MNRAS.423.3285V} and \citet{2013MNRAS.431..528J}. Table~\ref{table} shows the observed unsigned (large-scale) surface magnetic flux $\Phi_0$ (cf.~Eq.~(\ref{eq.phi0}) below) and the fractional energy in the poloidal axisymmetric modes $f_{\rm axi}$. The magnetic field that is initially considered in the grid is derived from extrapolations of observed surface radial magnetic maps using the potential-field source surface (PFSS) method \citep{1969SoPh....9..131A,2002MNRAS.333..339J}. The non-potential part of the observed field is not incorporated in our simulations, as it has been shown that stellar winds are largely unaffected by the non-potential, large-scale surface field \citep{2013MNRAS.431..528J}. The PFSS model assumes that the magnetic field is  potential ($\boldsymbol\nabla \times {\bf B}=0$) up to a radial distance $r=r_{\rm SS}$, which defines the source surface. Beyond $r_{\rm SS}$, all the magnetic field lines are considered to be open and purely radial, as a way to mimic the effects of a stellar wind. For all the cases studied here, we take $r_{\rm SS}=4~R_\star$, but we note that different values of $r_{\rm SS}$ produce the same final state solution for the simulations \citep{2011MNRAS.412..351V}. 

Once set at the initial state of the simulation, the values of the observed $B_r$ are held fixed at the surface of the star throughout the simulation run, as are the coronal base density and thermal pressure. A zero radial gradient is set to the remaining components of ${\bf B}$ and ${\bf u}=0$ in the frame corotating with the star. The outer boundaries at the edges of the grid have outflow conditions, i.e., a zero gradient is set to all the primary variables. The rotation axis of the star is aligned with the $z$-axis, and the star is assumed to rotate as a solid body. Our grid is Cartesian and extends in $x$, $y$, and $z$ from $-20$ to $20~R_\star$, with the star placed at the origin of the grid. BATS-R-US uses block adaptive mesh refinement (AMR), which allows for variation in numerical resolution within the computational domain. The finest resolved cells are located close to the star (for $r \lesssim 2~R_\star$), where the linear size of the cubic cell is $0.0097~R_\star$. The coarsest cell  has a linear size of $0.31~R_\star$ and is located at the outer edges of the grid. The total number of cells in our simulations is around $80$ million. As the simulations evolve in time, both the wind and magnetic field lines are allowed to interact with each other. The resultant solution, obtained self-consistently, is found when the system reaches steady state in the reference frame corotating with the star.  

\section{Simulation Results}\label{sec.results}
\subsection{Configuration of the embedded magnetic field}
\subsubsection{Alfv\'en surfaces}
\begin{figure*}
\includegraphics[width=85mm]{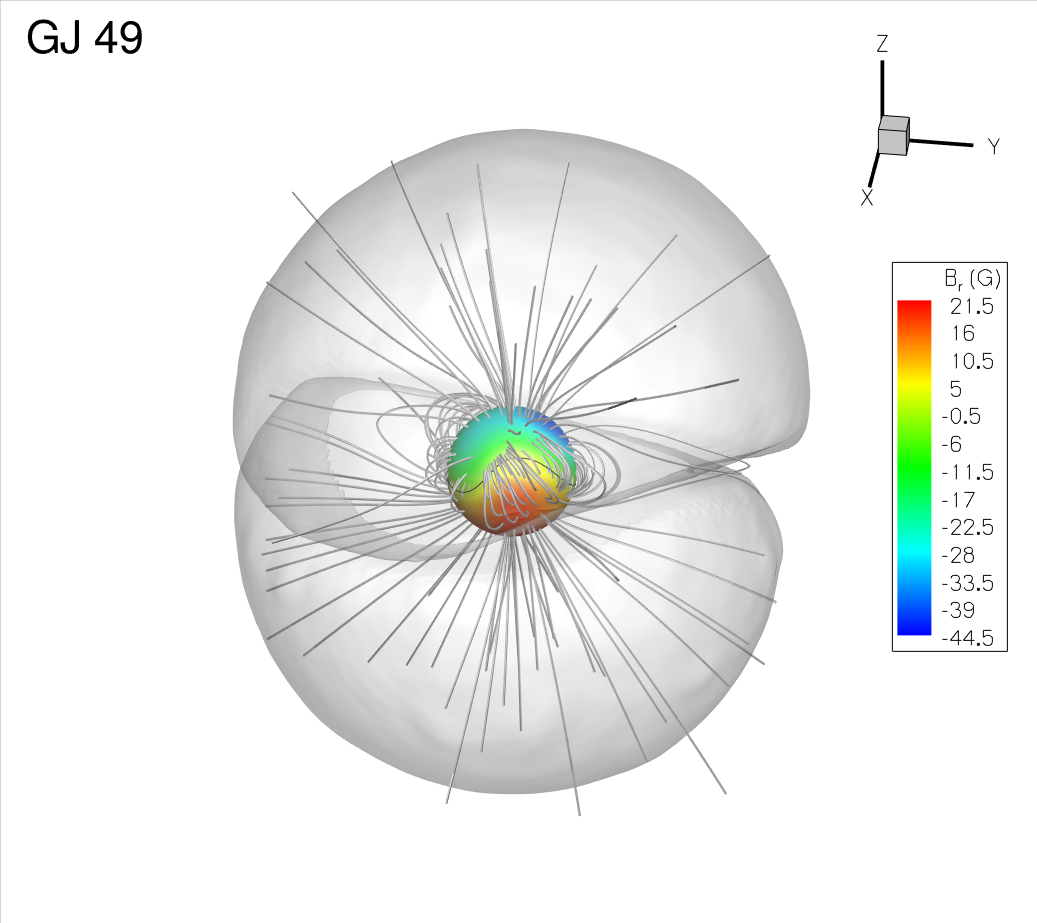}
\includegraphics[width=85mm]{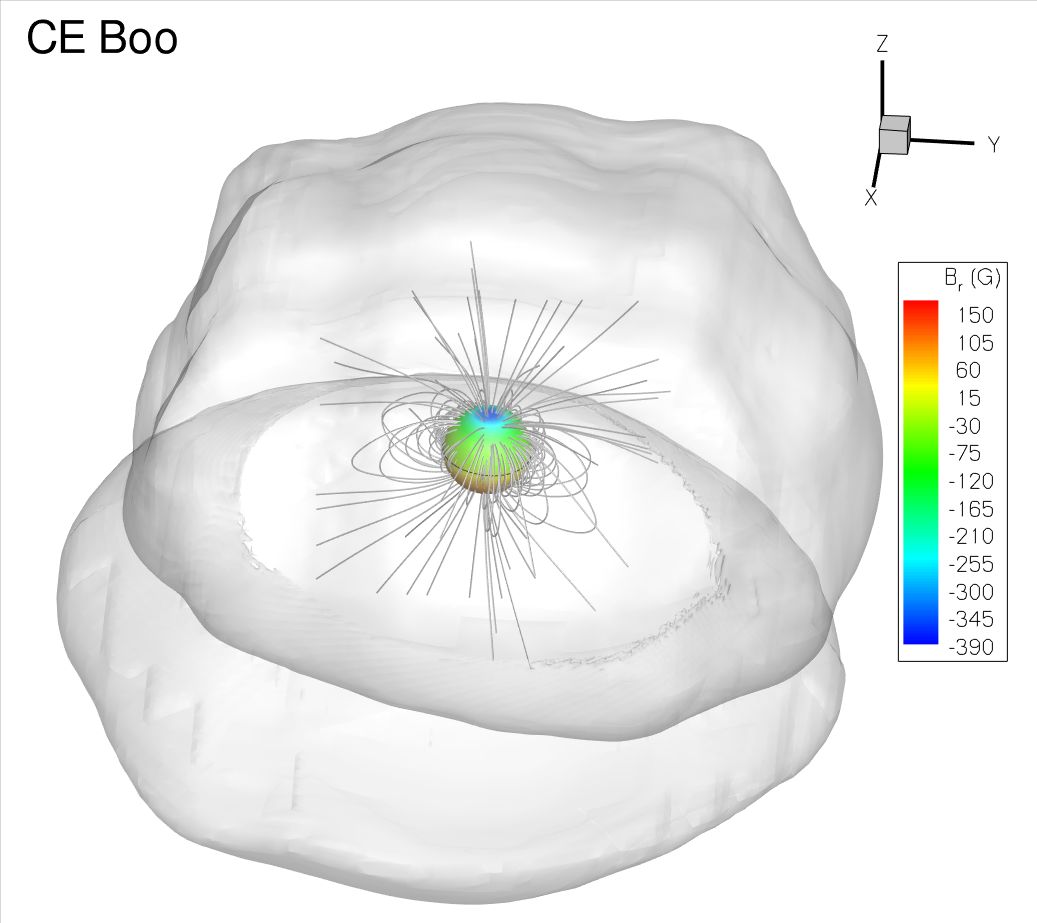}\\
\includegraphics[width=85mm]{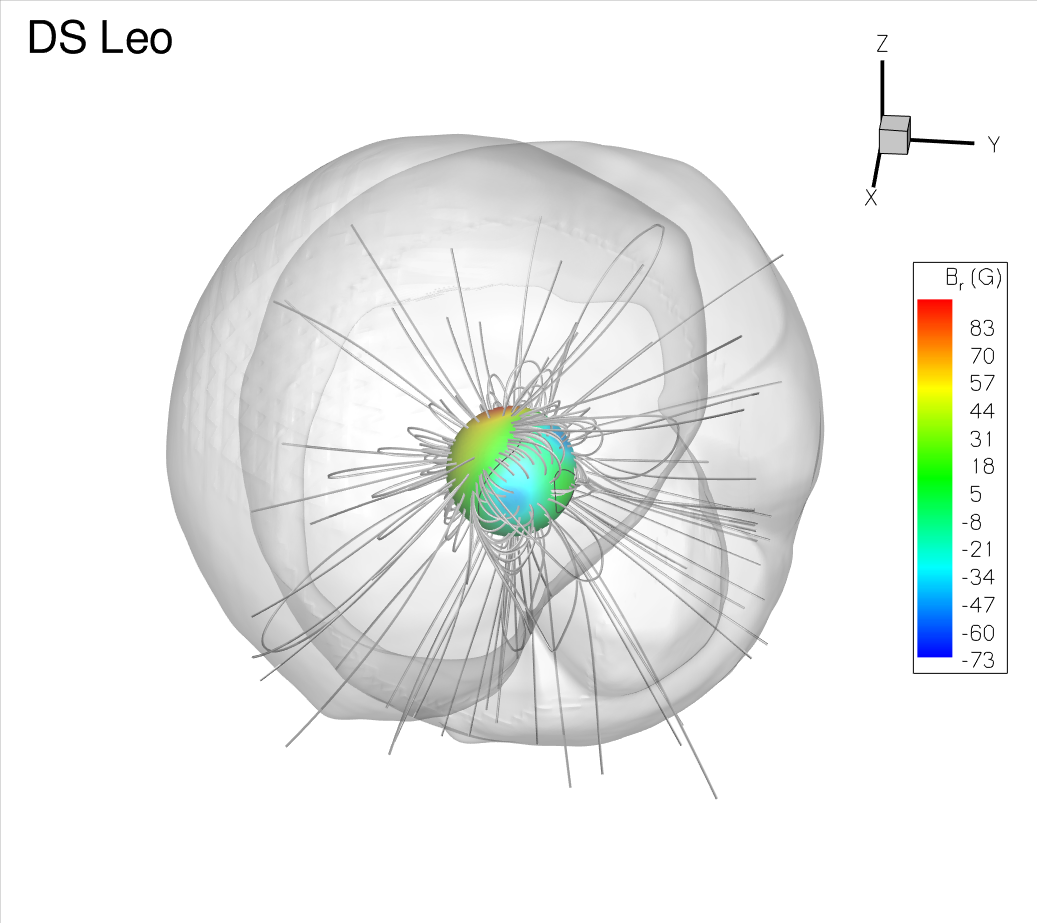}
\includegraphics[width=85mm]{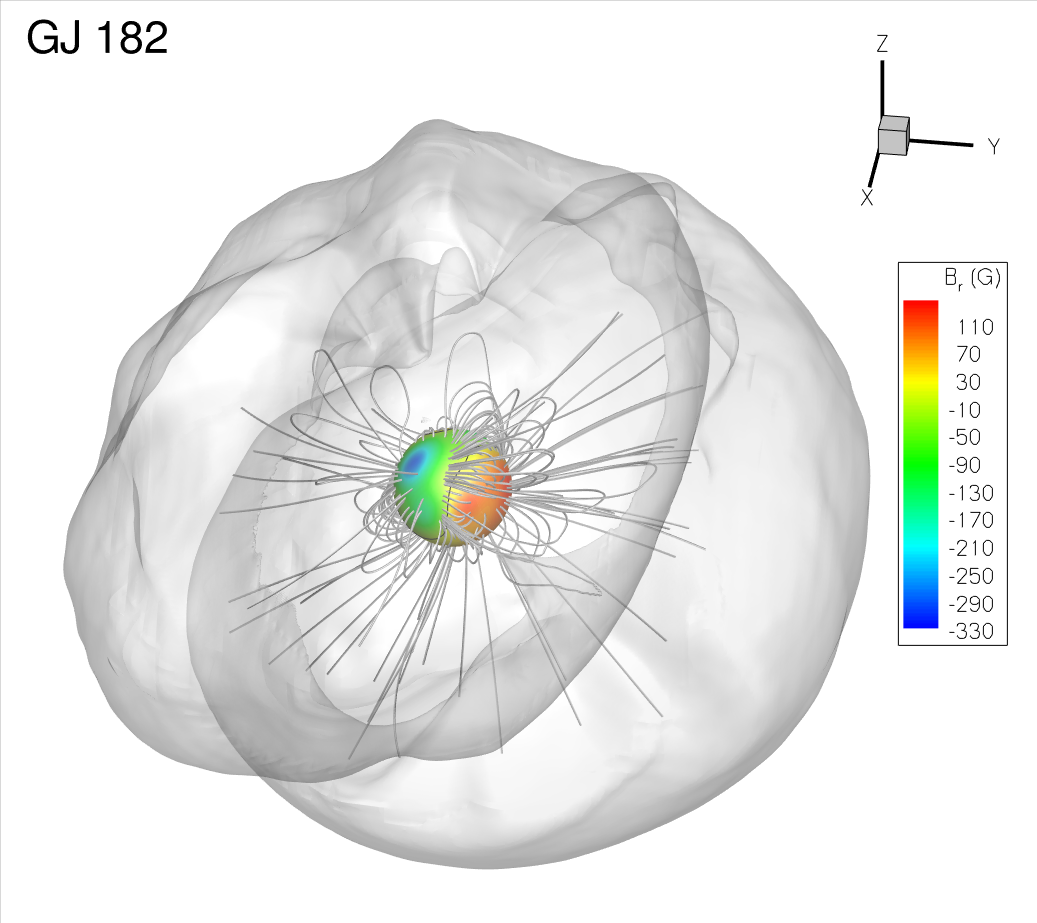}\\
\includegraphics[width=85mm]{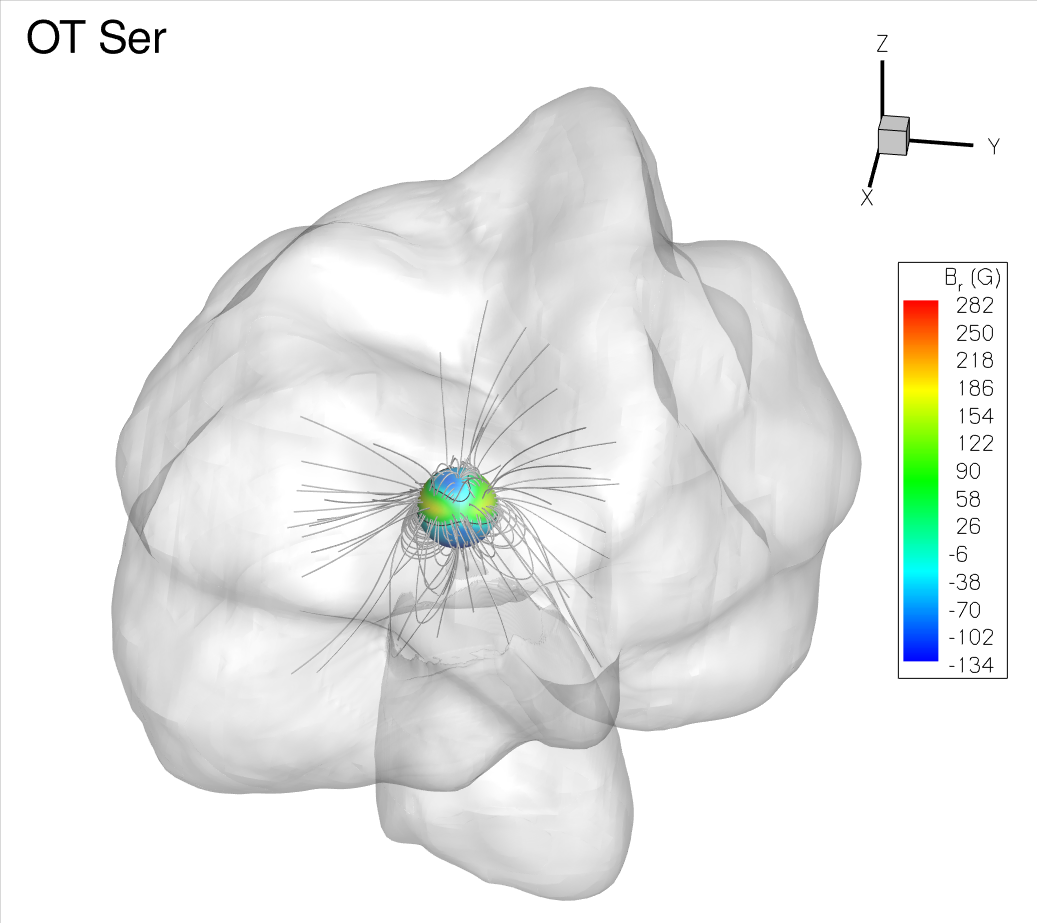}
\includegraphics[width=85mm]{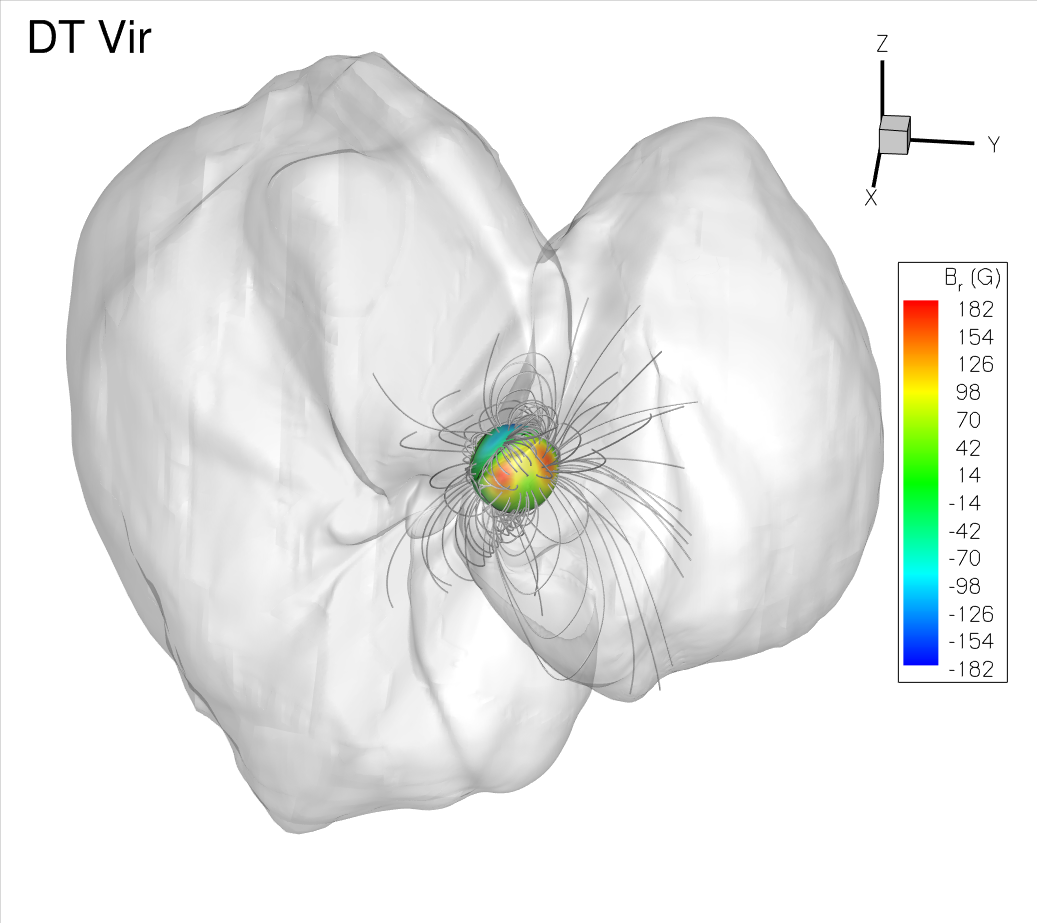}
\\
\caption{The final configuration of the magnetic field lines after the wind solution has relaxed in the grid. Over-plotted at the surface of the star is the observationally reconstructed stellar magnetic field \citep{2008MNRAS.390..545D}, used as boundary condition for the radial magnetic field. \alf\ surfaces are shown in grey. Note their irregular, asymmetric shapes due to the irregular distribution of the observed field.\label{fig.IC-SS}}
\end{figure*}

Figure~\ref{fig.IC-SS} shows the final configuration of the magnetic field lines obtained through self-consistent interaction between magnetic and wind forces after the simulations reached steady state. Although we assume the magnetic field is potential in the initial state of our simulations, this configuration is deformed when the interaction of the wind particles with the magnetic field lines (and vice-versa) takes place. Figure~\ref{fig.IC-SS} also shows the Alfv\'en surface $S_A$ in grey. This surface is defined as the location where the wind velocity reaches the local Alfv\'en velocity ($v_A = B(4 \pi \rho)^{-1/2}$). Inside $S_A$, where the magnetic forces dominate over the wind inertia, the stellar wind particles are forced to follow the magnetic field lines. Beyond $S_A$, the wind inertia dominates over the magnetic forces and, as a consequence, the magnetic field lines are dragged by the stellar wind. In models of stellar winds, the \alf\ surface has an important property for the characterisation of angular momentum losses, as it defines the lever arm of the torque that the wind exerts on the star (cf.~Eq.~(\ref{ap.eq.jdot_sa})). Contrary to the results obtained on wind models with axisymmetric magnetic fields, the \alf\ surfaces of the objects investigated here have irregular, asymmetric shapes, which can only be captured by fully three-dimensional wind models. Note that these odd shapes are consequence of the irregular distribution of the observed magnetic field.

\subsubsection{Effective source surface}
The PFSS method has proven to be a fast and simple way to extrapolate surface magnetic fields into the stellar coronal region \citep{1999MNRAS.305L..35J,2002MNRAS.333..339J, 2013A&A...557A..67V}. It is also used here as the initial conditions for our simulations. The free parameter of the PFSS method is the radius $r_{\rm SS}$ of the source surface, beyond which the magnetic field lines are assumed open and radial (Section~\ref{sec.model}). To constrain values of $r_{\rm SS}$ to be used by PFSS methods, we wish to provide here, from our fully 3D MHD models, an effective radius of the source surface. Motivated by the approach used in \citet{2006ApJ...653.1510R}, we define an `effective source surface' ($r_{\rm SS}^{\rm eff}$) for the MHD models as the radius of the spherical surface where $97$~percent of the average magnetic field is contained in the radial component (i.e., $\langle |B_r| \rangle / \langle |B| \rangle =0.97$). 
For the fastest rotating stars in our sample (DT~Vir, OT~Ser, GJ~182), the ratio $\langle |B_r| \rangle / \langle |B| \rangle$ does not reach the 97-percent level, as in these cases the $B_\varphi$ contribution has a relatively larger weight. In such cases, we take $r_{\rm SS}^{\rm eff}$ to be the position where $\langle |B_r| \rangle / \langle |B| \rangle$ is maximum. Table~\ref{table2} shows the derived values of $r_{\rm SS}^{\rm eff}$. For the sample of stars analysed in this paper, we find that on average $r_{\rm SS}^{\rm eff} \simeq 3.65 \pm 0.77~R_\star$, indicating a  compact region of closed field lines. We note that this size is similar to the  usual adopted size of $2.5~R_\odot$ from PFSS methods of the solar coronal magnetic field.

\begin{table*} 
\centering
\caption{Derived values from the simulations. The columns are, respectively: the star name, the unsigned open magnetic fluxes ($\Phi_{\rm open}$),  the effective radius of the source surface derived from the MHD models ($r_{\rm ss}^{\rm eff}$), mass-loss rate contrast ratio ($f_{\dot{M}}= \max (\rho u_r)/\min(\rho u_r)$) calculated at a distance $r\simeq 19~R_\star$ (near the edge of our simulation domain), the angular momentum-loss rate calculated from our 3D simulations ($\dot{J}$), the ratio between our derived $\dot{J}$ and from a simplified 1D model ($\dot{J}/\dot{J}_{\rm 1D}$), the instantaneous spin-down time ($\tau$), and the estimated age of the star (see text). \label{table2} }    
\begin{tabular}{lccccccccccccccccc}  
\hline
Star	&	$	\Phi_{\rm open}	$		&$r_{\rm ss}^{\rm eff}$	&	$	f_{\dot{M}}				$				&	$\dot{J}$&$\dot{J}/\dot{J}_{\rm 1D}$			&	$	\tau		$	&		Age		\\		
ID			&	$	(\Phi_{0})	$ & $(R_\star)$ &		&$(10^{32}$ erg)& &		(Myr)			&		(Myr)	\\
 \hline	
    GJ 49  & $ 0.56$ & $  2.8$ & $  1.2$ & $ 0.39$ & $ 0.45$ & $ 1816$ & $ 1190 $ \\
    CE Boo  & $ 0.46$ & $  4.6$ & $  2.4$ & $  1.4$ & $ 0.15$ & $  396$ & $  125 $ \\
    DS Leo  & $ 0.45$ & $  3.6$ & $  1.4$ & $ 0.96$ & $ 0.50$ & $ 1037$ & $  710 $ \\
    GJ 182  & $ 0.50$ & $  4.6$ & $  4.1$ & $   85$ & $ 0.36$ & $  120$ & $   12 $ \\
    OT Ser  & $ 0.48$ & $  3.0$ & $  3.4$ & $   12$ & $ 0.20$ & $  292$ & $   - $ \\
    DT Vir  & $ 0.30$ & $  3.6$ & $  2.8$ & $   16$ & $ 0.39$ & $  328$ & $   - $ \\
 \hline			
\end{tabular}
\end{table*}

\subsection{Derived properties of the stellar winds}

\begin{figure*}
\includegraphics[height=60mm]{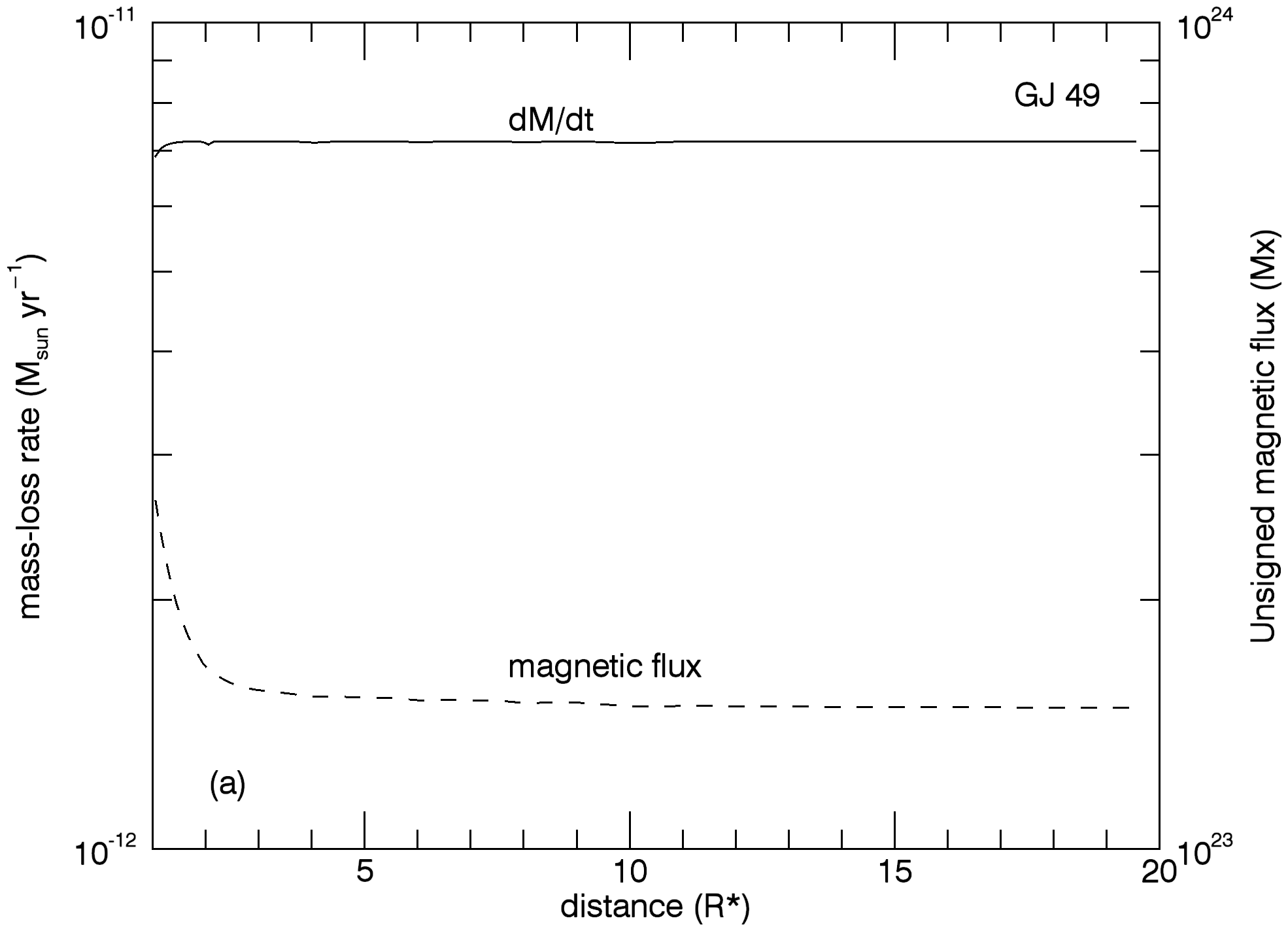}
\includegraphics[height=60mm]{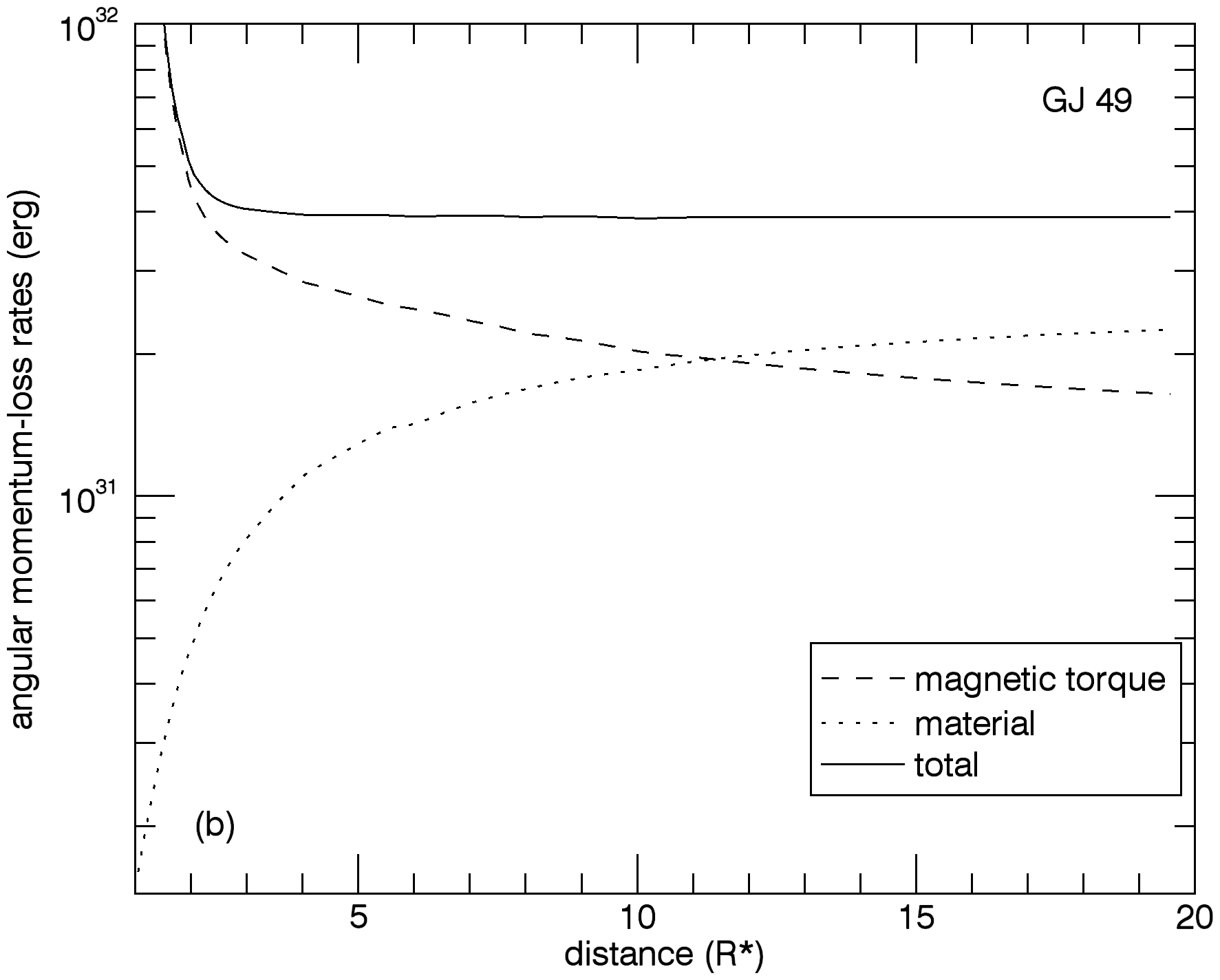}
\caption{Radial dependence of (a) mass-loss rate (solid line) and unsigned magnetic flux (dashed line) and (b) total angular momentum-loss rate (solid line) for GJ~49, illustrating conservation of these quantities in our simulations. Panel (b) also shows how the magnetic torque (dashed line) and the angular momentum of the material (dotted line) contribute to $\jdot$ (Eq.~(\ref{eq.jdot})). \label{fig.conservation}}
\end{figure*}

Table~\ref{table2} presents the properties of the stellar wind derived from our simulations. The unsigned observed surface magnetic flux is defined as
\begin{equation}\label{eq.phi0}
\Phi_0 = \oint_{S_\star} |B_r (R_\star, \theta, \varphi)| {\rm d} S_\star
\end{equation}
(Table~\ref{table}) and the unsigned open magnetic flux as
\begin{equation} \label{eq.phiopen}
\Phi_{\rm open} = \oint_{S_{\rm sph}} |B_r (r, \theta, \varphi)| {\rm d} S_{\rm sph}. 
\end{equation}
The former is integrated over the surface of the star $S_\star$ and the latter over a spherical surface $S_{\rm sph}$ at a large distance $r$ from the star, where all the field lines are open. Figure~\ref{fig.conservation}a shows the unsigned magnetic flux (dashed line) as a function of distance for the simulation of GJ~49. Note that for large distances, the flux is $ \Phi_{\rm open}$ and is conserved in our simulation. In our simulations, magnetic fluxes are conserved within $1.5$ per cent.

\subsubsection{Mass flux}\label{sec.massflux}
\begin{figure*}
\includegraphics[width=56mm]{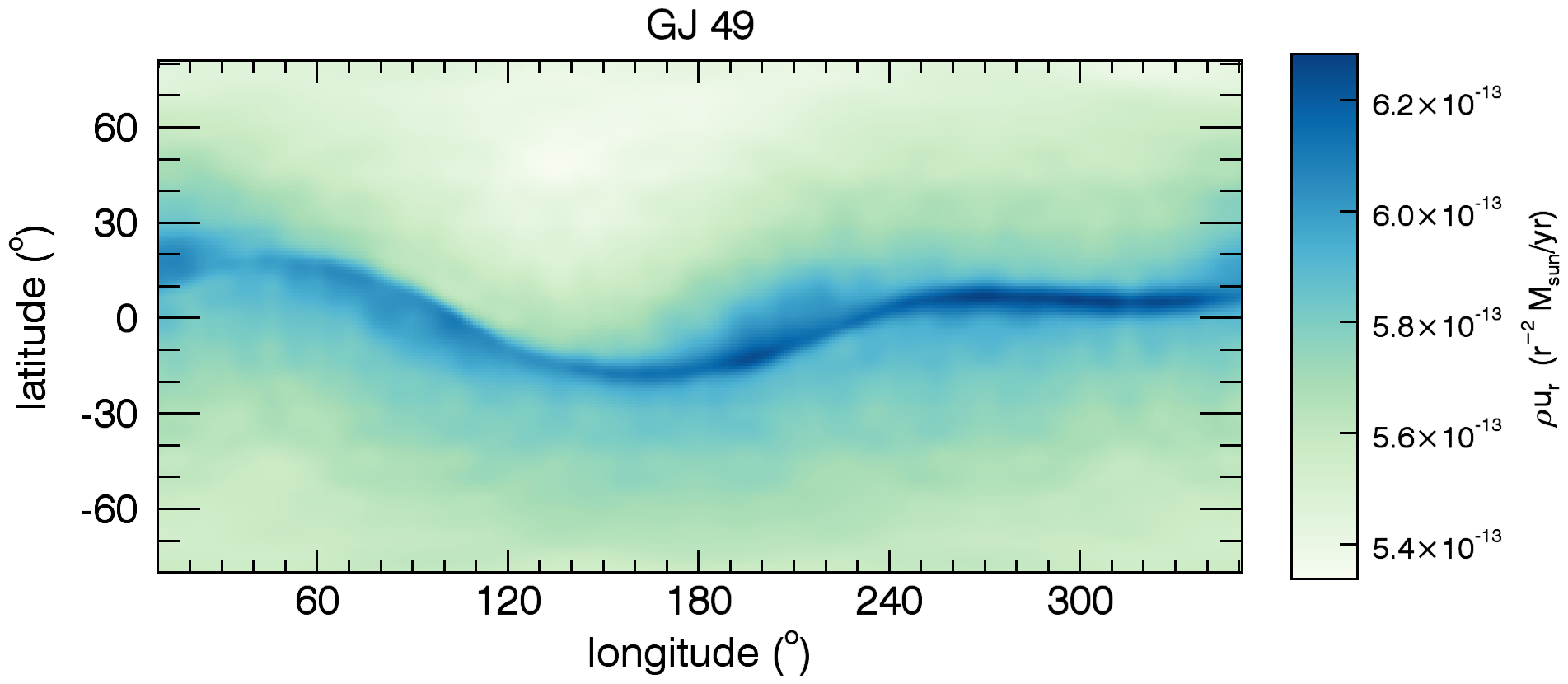}
\includegraphics[width=56mm]{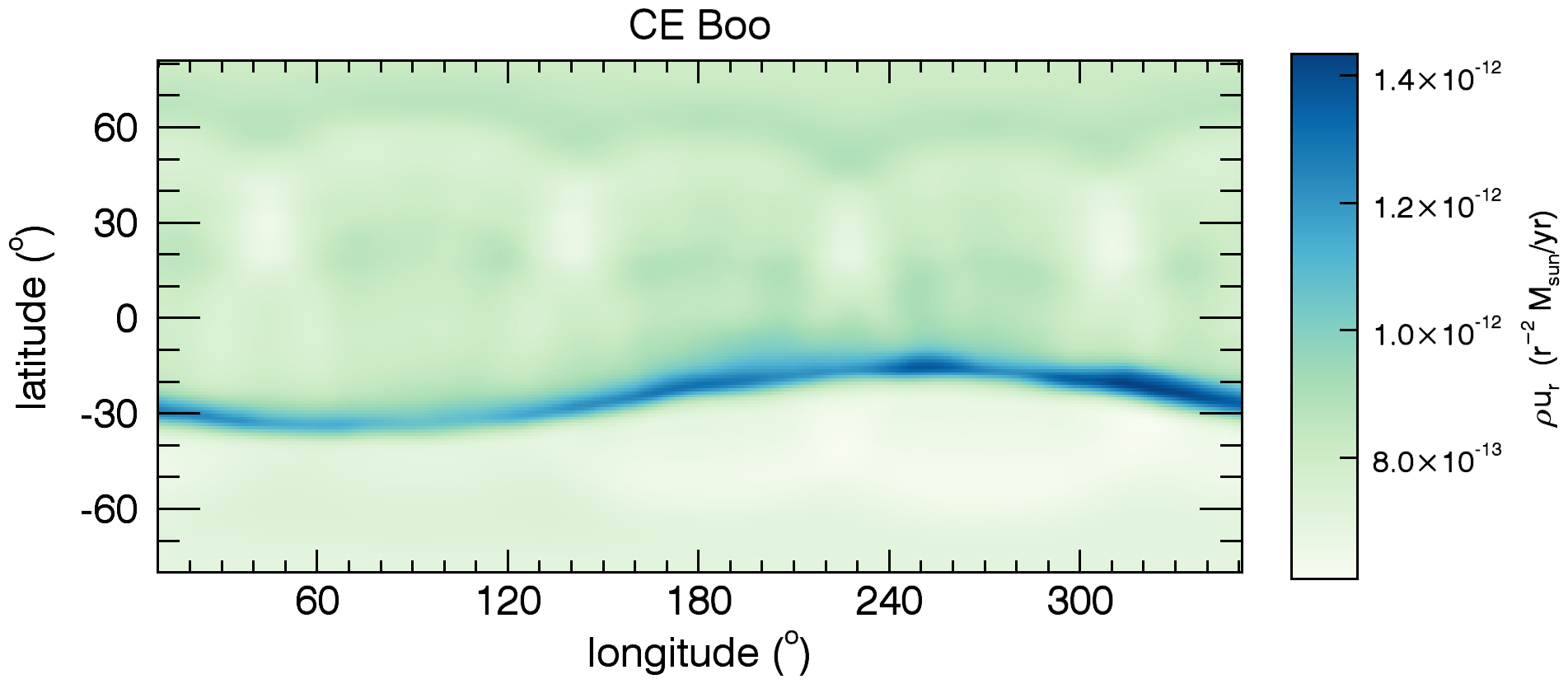}
\includegraphics[width=56mm]{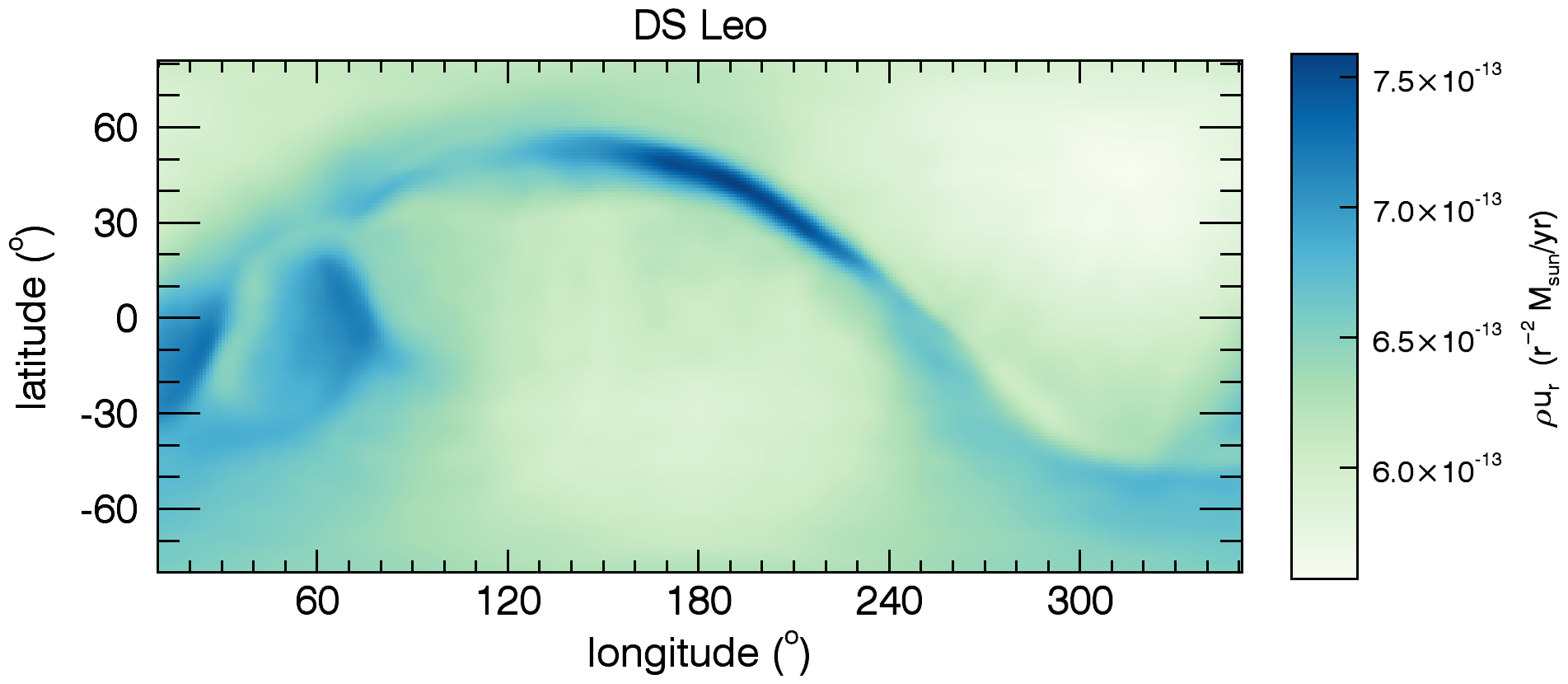}\\
\includegraphics[width=56mm]{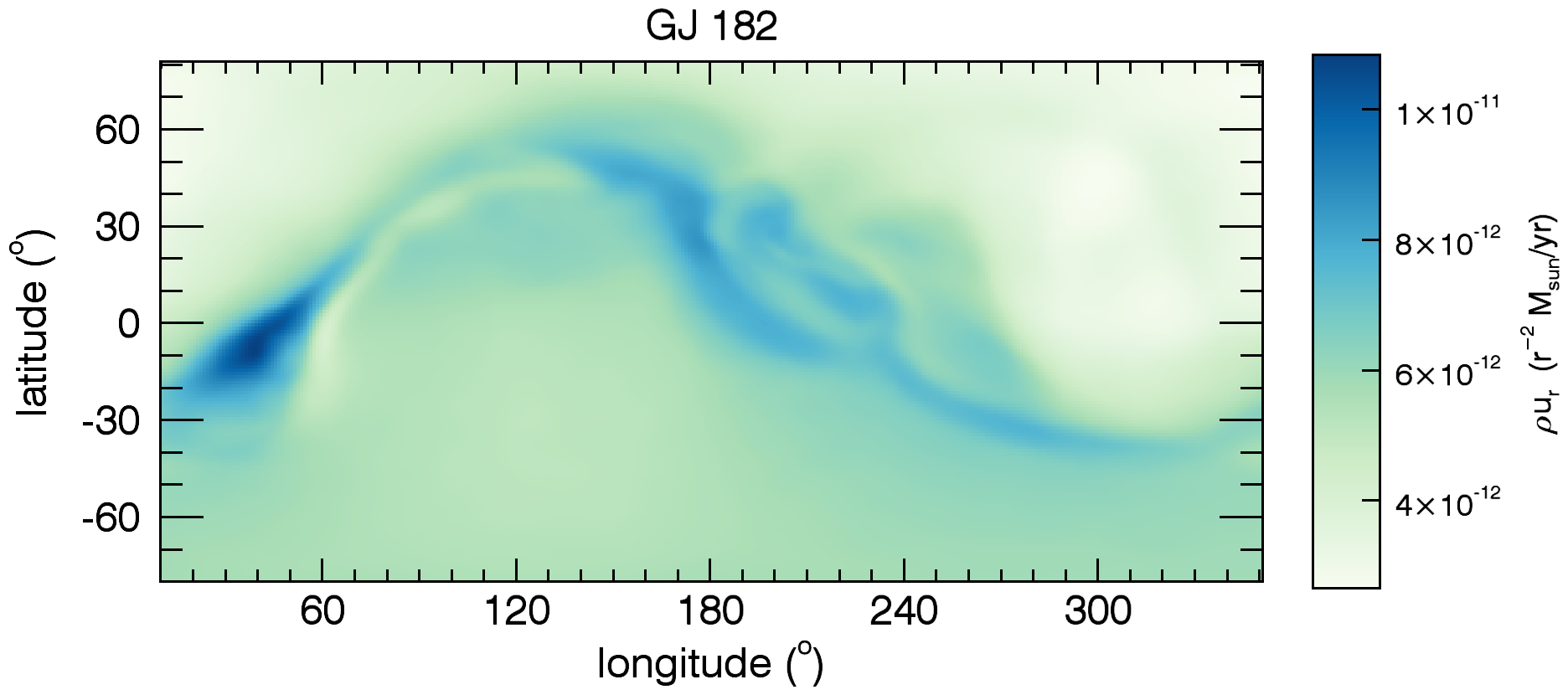}
\includegraphics[width=56mm]{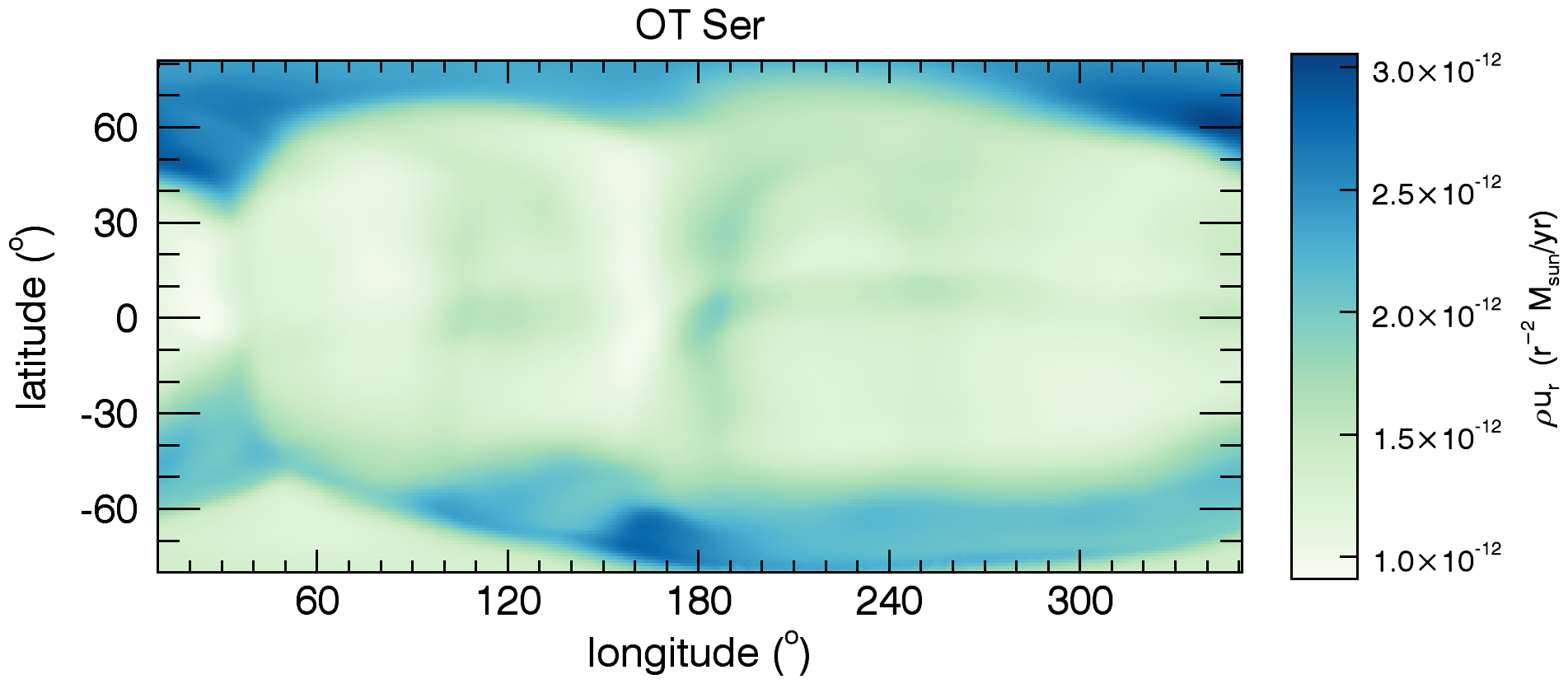}
\includegraphics[width=56mm]{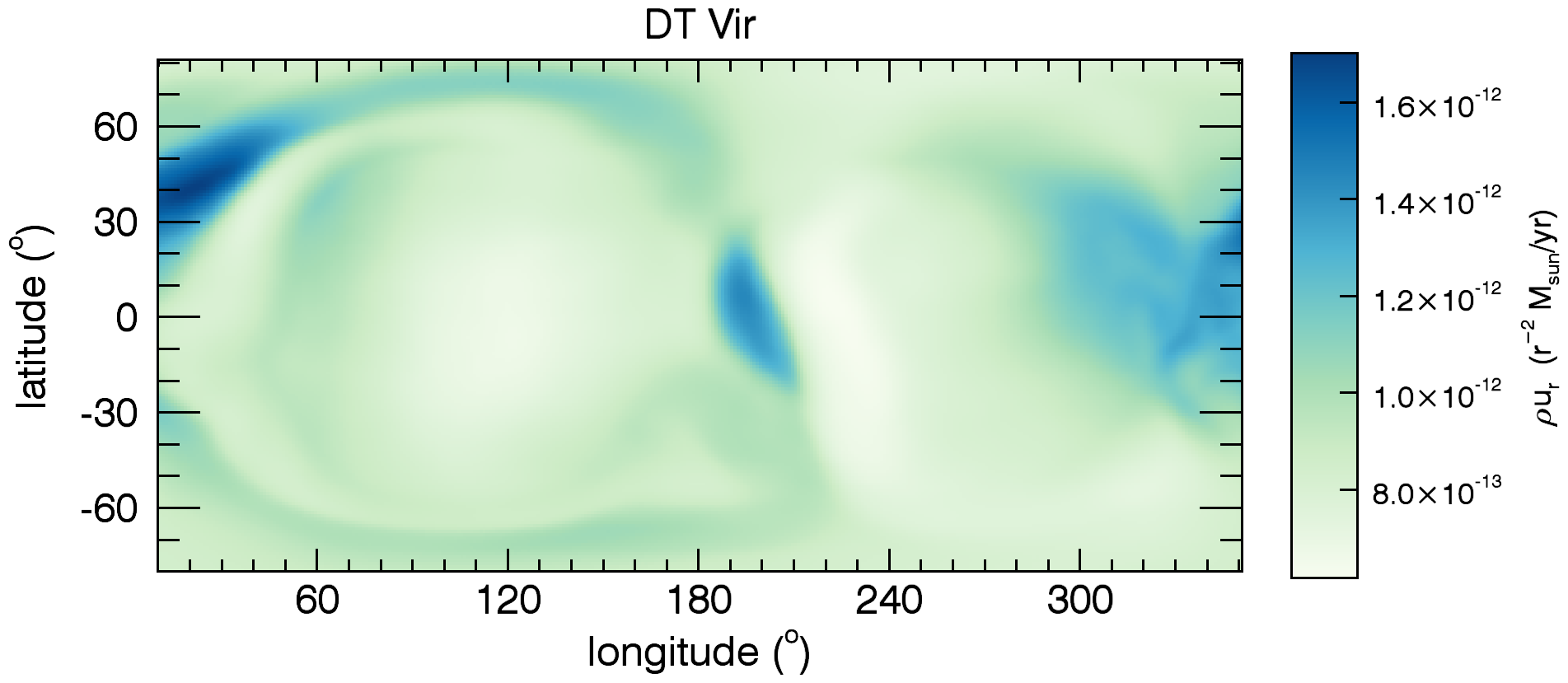}\\
\caption{Distribution of mass flux $\rho u_r$ across a spherical surface of radius $r \simeq 19~R_\star$ (close to the edge of our simulation domain). The more asymmetric topology of the stellar magnetic field results in more asymmetric mass fluxes. \label{fig.massflux}}
\end{figure*}

The mass-loss rate $\dot{M}$ is defined as the flux of mass integrated across a closed surface $S$ 
\begin{equation}\label{eq.mdot}
\dot{M} = \oint_S \rho {\bf u} \cdot {\rm d} {\bf S} = \oint \rho u_r {\rm d} S_{\rm sph},
\end{equation}
where $\dot{M}$ is a constant of the wind. Figure~\ref{fig.conservation}a shows the mass loss-rate (solid line) as a function of distance for the simulation of GJ~49. In our simulations, $\mdot$ are conserved within $0.2$ per cent at most. Figure~\ref{fig.massflux} shows the distribution of mass flux $\rho u_r$ across a spherical surface of radius $\sim 19~R_\star$ (close to the edge of our simulation domain). As can be seen, the mass flux is not homogeneously distributed, as would be the case of a spherically symmetric wind. At $r\simeq19~R_\star$, the mass flux has a contrast ratio $f_{\dot{M}}= \max (\rho u_r)/\min(\rho u_r)$ of a factor of up to $\sim 4$ (see Table~\ref{table2}), but we note that at closer distances to the star, the distribution of mass flux is different. Mass fluxes are therefore redistributed latitudinally with distances by meridional flows. As our simulation domain extends only out to $20~R_\star (\sim 0.04~{\rm au})$, we do not know if this trend is kept for larger $r$. In the case of the solar wind, \citet{2011MNRAS.417.2592C} finds, from in situ measurements of the solar wind taken by WIND/ACE (near $1$~au) and by Ulysses (at high heliographic latitudes between $3$ and $5$~au), that the solar mass-loss rate is roughly the same at different latitudes for large distances. Note also that  the mass-loss rate of the solar wind has a spread of more than one order of magnitude (Figures 1 and 3 from \citealt{2011MNRAS.417.2592C}), which could mask smaller variations predicted in our theoretical studies.

In addition, we note that the mass flux profile is essentially modulated by the local value of $|B_r|$, which share roughly the same characteristics as the surface $|B_r|$. Therefore, we conclude that the more non-axisymmetric topology of the stellar magnetic field results in more asymmetric mass fluxes. For example, if we were to place a spacecraft that can only provide mass flux measurements at a single location of the wind, such a spacecraft would estimate total mass-loss rates incorrectly, because it would neglect longitudinal and latitudinal differences. In addition, latitudinal and longitudinal variations in mass flux should also affect the distances and shapes of astropauses\footnote{In analogy to the heliopause, the astropause is defined as the surface where the pressure of the stellar wind balances the pressure of the interstellar medium.}, which would lack symmetry due to the non-axisymmetric nature of the stellar magnetic field. 

\subsubsection{Angular momentum flux}
The outflow per unit area of the $z$-component of the angular momentum flux across a closed spherical surface $S_{\rm sph}$ is 
\begin{equation}\label{eq.jdot}
\dot{J} = \oint_{S_{\rm sph}} \left[  - \frac{\varpi B_\varphi B_r}{4 \pi} + \varpi u_\varphi \rho u_r \right]  {\rm d} S _{\rm sph} 
 \end{equation} 
where $\varpi=(x^2+y^2)^{1/2}$ is the cylindrical radius. Appendix~\ref{ap.amloss} shows the derivation of Eq.~(\ref{eq.jdot}), which follows from the derivation presented in \citet{1970MNRAS.149..197M} and \citet{1999stma.book.....M}. We find that angular momentum-loss rates range between $10^{31}$ to almost $10^{34}$~erg for the stars in our sample\footnote{The wind base density is a free parameter of our model, which could be constrained by more precise measurements of mass-loss rates of dM stars (cf. Section~\ref{sec.model}). Our models assume a wind number density of $n_0=10^{11}$~cm$^{-3}$. In order to investigate how a different choice of $n_0$ would affect our derived mass and angular momentum loss rates, we performed a simulation for one of the cases studied (GJ~49) in which we adopted a different base density ($n_0=10^{10}$~cm$^{-3}$). We found a decrease in loss rates (a factor of $0.12$ and $0.43$ in mass- and angular momentum-loss rates, respectively) as compared to values of the simulation where a larger density was adopted.}. Figure~\ref{fig.conservation}b shows the angular momentum loss-rate as a function of distance for the simulation of GJ~49. In our simulations, $\jdot$ are conserved within $5$ per cent at most.

Table~\ref{table2} also presents an estimate of the instantaneous time scale for rotational braking, defined as $\tau = {J}/{\dot{J}}$, where $J$ is the angular momentum of the star. If we assume a spherical star with a uniform density, rotating at a rate $\Omega_\star$, then $J = \frac{2}{5} M_\star R_\star^2 \Omega_\star$ and the time scale is estimated as
\begin{equation}
\tau \simeq \frac{8.9 \times 10^{36}}{\dot{J} {\rm [erg]}} \left( \frac{M_\star}{M_\odot}\right) \left( \frac{1~{\rm d}}{P_{\rm rot}} \right) \left(  \frac{R_\star}{R_\odot}\right)^2 ~{\rm Myr}.
\end{equation} 
The constant in the equation above ($8.9 \times 10^{36}$~Myr) equals $\frac25 M_\odot R_\odot^2 (2\pi/P_{\rm rot})$, for $P_{\rm rot}=1$~d. The values of $\tau$ obtained here are representative of the epoch when the magnetic surface maps were derived and it is likely that they vary with the evolution of the magnetic field topology of the star.  In the Sun, for example, angular momentum loss may be enhanced in certain phases of the stellar cycle, alternating between epochs with a greater and smaller releases of angular momentum \citep{2011ApJ...737...72P}. Because we do not know if the stars analysed here present a magnetic cycle, we do not know if they will present a cyclic variation in $\dot{J}$, similar to the solar case. For that, a long-term monitoring of these stars would be required.

Table~\ref{table2} shows the estimated age for some of the objects. Ages for GJ 182 and CE Boo are more reliable, as the first one is part of the $\beta$-Pic association ($\sim 12$~Myr, \citealt{2006A&A...460..695T}) and the latter is a member of the Pleiades open cluster ($\sim 125$~Myr, \citealt{1998ApJ...499L.199S}). Observations suggest that at about the age of $625~$Myr, main-sequence early-dM stars should have spun down to a tight colour-period relation \citep{2011MNRAS.413.2218D}. Using the colour-period relation derived by \citet{2011MNRAS.413.2218D}, we obtain age estimates for GJ 49 and DS Leo. The gyrochronology method can not be reliably applied for early-dM stars with $P \lesssim 13~$d, such as DT Vir and OT Ser, as these objects may not have converged towards the tight colour-period sequence. We note that, whenever available, the estimated ages suggest that the objects in our sample seem to be much younger than the Sun, a consequence of the faster rotation of the former. 

For the cases with age estimates, we see that the instantaneous spin-down time scales exceed the ages, suggesting that the stars in our samples may not have had enough time to spin down. We note that all the stars studied here are active ones, which are the most accessible to ZDI studies. It is expected that, as the star spins down, the efficiency in producing strong magnetic fields is reduced, resulting in old objects that are less active and slowly rotating. The time scale for that to happen depends on the mass of the objects (cf. Section~\ref{sec.intro}).

\subsubsection{Meridional structure of mass and angular momentum fluxes}
\begin{figure}
\includegraphics[width=80mm]{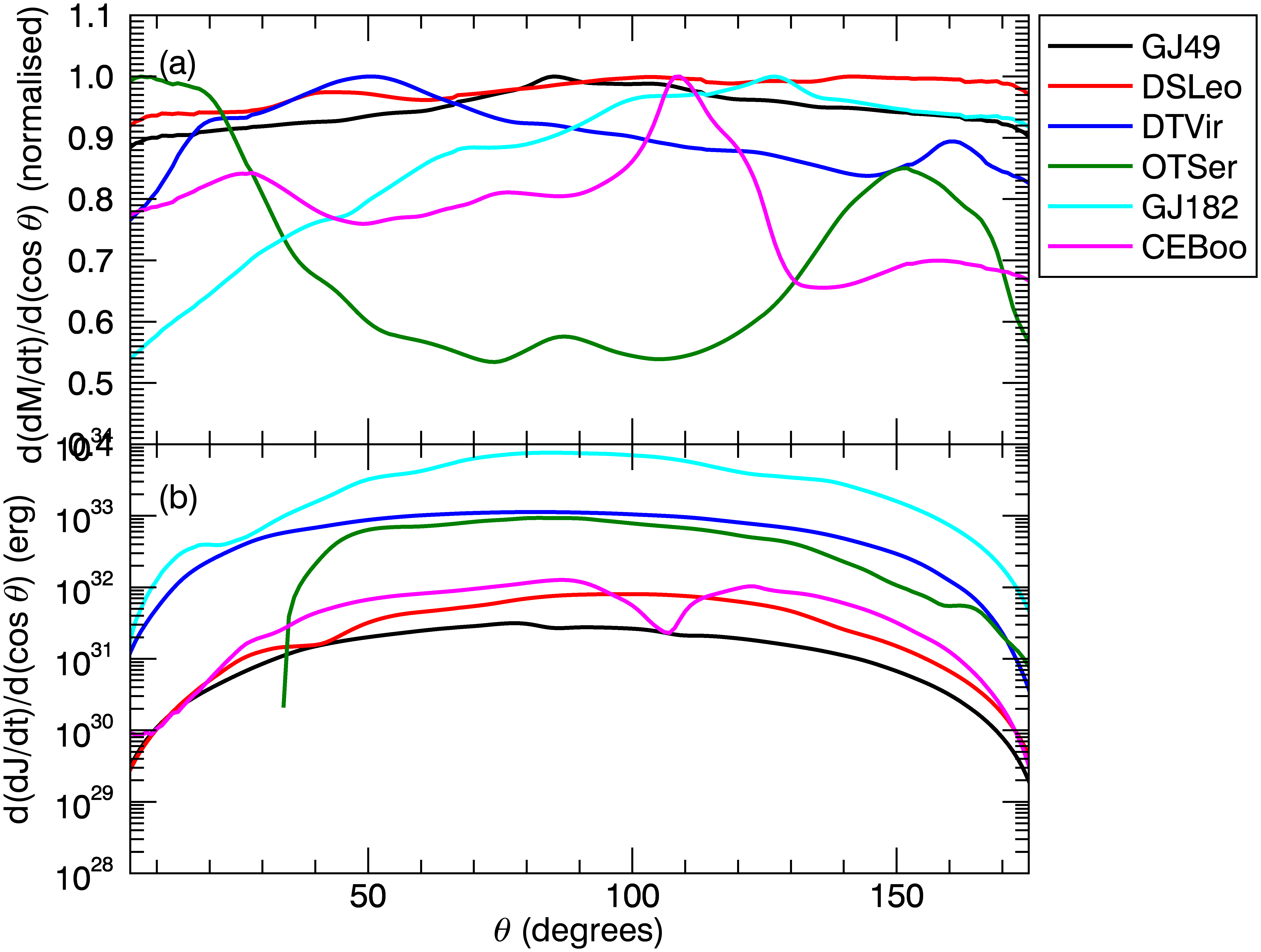}
\caption{Meridional distribution of (a) mass- and (b) angular momentum-loss rates calculated close to the outer edge of our grid ($r \simeq 19~R_\star$).  \label{fig.meridional}}
\end{figure}

Figure~\ref{fig.meridional} shows how mass- and angular momentum-loss rates vary as a function of colatitude $\theta$ for the six stars studied in this paper. To compute that, we integrate mass and angular momentum fluxes along azimuth ($\varphi$). By doing that, we lose information on the non-axisymmetric features of each of these fluxes. On the other hand, these integrations allow us to investigate which colatitude on average contributes more to angular momentum- and mass-loss rates.   

For the mass-loss rate (Fig.~\ref{fig.meridional}a, normalised to the maximum value of $d \mdot/ d\cos\theta$ for each object), most of the asymmetric variability seen in Figure~\ref{fig.massflux} is washed out, as seen by the almost `flat' profiles of mass flux versus $\theta$. We find that there is no preferred colatitude that contributes more to mass loss, as the mass flux is maximum at different colatitudes for different stars.
  
For the angular momentum flux (Fig.~\ref{fig.meridional}b), even after azimuthal integration, there is still a significant contrast of a few orders of magnitude between regions near the poles and the equator.  At the poles, the angular momentum flux goes to zero (because $\varpi \to 0$ in Eq.~(\ref{eq.mdot})), while it is maximum in regions within $\sim 10^\circ$ above and below the equator (at $\theta=78^\circ$ for GJ~49 and at $\theta=97^\circ$ for DS~Leo). This indicates that the flow at equatorial regions carries most of the stellar angular momentum. We also find that, at different distances, the same characteristic of angular momentum flux as a function of $\theta$ persists ($\jdot$ is not redistributed over colatitudes). Note however that the individual contributions (magnetic or kinetic, in Eq.~(\ref{eq.jdot})) to the angular momentum transport have their $\theta$-profiles altered at different distances, as one type of transport is converted to the other. Note that the azimuthally integrated profiles shown in Figure~\ref{fig.meridional}b are qualitatively similar to the one obtained by \citet{1993MNRAS.262..936W}, who considered an axisymmetric dipolar magnetic field distribution (see their Figure~3).  

\subsection{Dependences with observables}\label{sec.fits_obs}
\begin{figure}
\includegraphics[width=80mm]{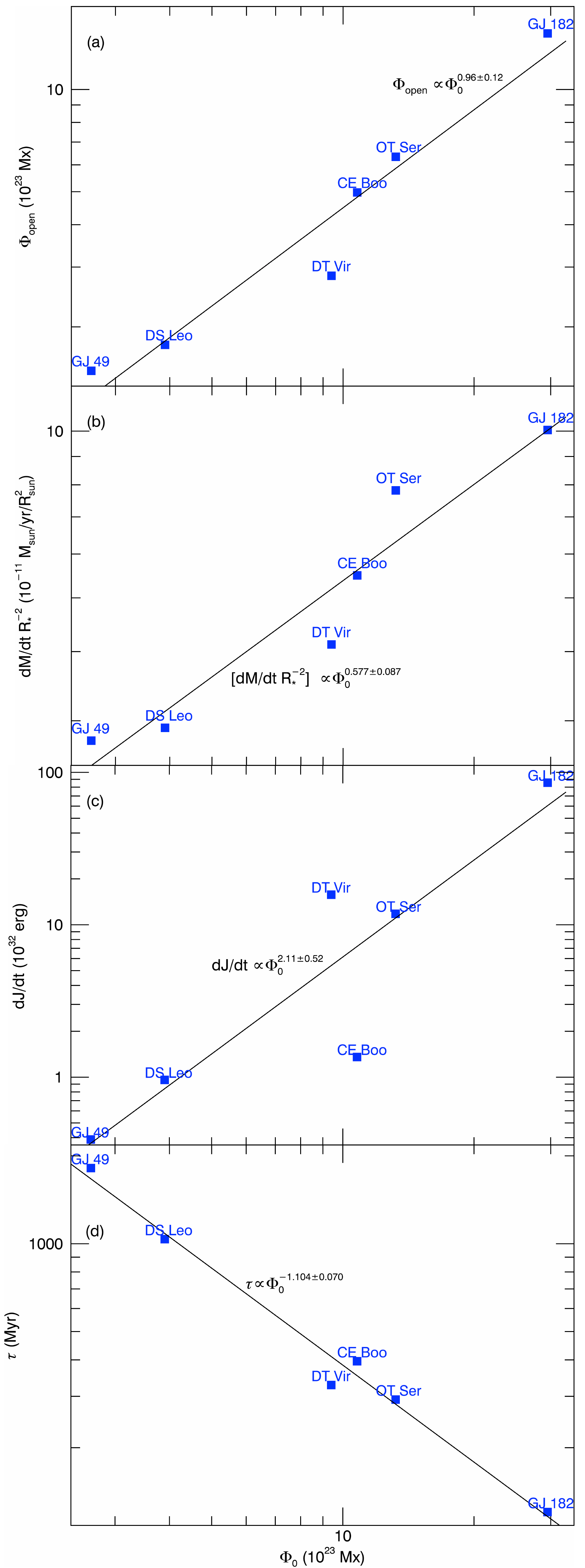}
\caption{Dependence of a few selected wind-derived parameters with the unsigned surface magnetic field $\Phi_0$. Solid lines are power-law fits of the type $f(\Phi_0) \propto \Phi_0^{p}$, where the dependent variable $f$ is, from top to bottom, $\Phi_{\rm open}$, $\mdot/R_\star^2$, $\jdot$ and $\tau$, respectively. The power-law indexes $p$ of the fits are shown in the top portion of Table~{\ref{tab.fits}}.  \label{fig.fit_obs}}
\end{figure}

In this section, we provide relations between the output of our simulations with observable quantities. Our goal is to provide a fast method to estimate stellar wind quantities once observable parameters become available. The results of our simulations were plotted against observed quantities (and against themselves, as will be presented in Section~\ref{sec.modelcomp}) and we then fitted power laws of the type $f(x) \propto x^{p}$, where the dependent variable is $f$, the independent variable is $x$, and $p$ is the power-law index derived from the fitting procedure (Figure~\ref{fig.fit_obs}). The errors computed for $p$ are associated to the fit procedure only. Numerical errors or observable errors were not considered in our fits. We note that the sample provided here is small, containing only six early-dM stars, so that a statistical analysis is out of reach. In addition, these stars present different magnetic field properties, which in most of the cases results in poor fits. A larger set of simulations should be carried out in order to confirm the relations we found. The top portion of Table~{\ref{tab.fits}} presents a few selected power-law fits. 

\begin{table} 
\centering
\caption{Selected relations between the output of our simulations with the observable unsigned magnetic surface flux (top portion of the table) and with wind derived parameters (bottom). The exponents of the relations were derived from power-law fits. Numerical errors or observable errors were not considered in our fits. Errors displayed below are associated to the fit procedure only. 
\label{tab.fits}}    
\begin{tabular}{llccccccccccccccc}  
 \hline	
$\Phi_{\rm open} \propto \Phi_0^{0.96 \pm 0.12}$\\
$\mdot/R_\star^2 \propto \Phi_{0}^{0.577\pm 0.087} $\\
$\jdot \propto \Phi_0 ^{2.11 \pm 0.52} $\\
$\tau \propto \Phi_0 ^{ -1.104 \pm 0.070} $\\ \hline
$\mdot \propto \Phi_{\rm open}  ^{ 0.89\pm 0.19 }$\\
$\jdot\propto \Phi_{\rm open}^{1.96 \pm 0.68 }$\\
$\mdot/R_\star^2\propto \Phi_{\rm open}^{ 0.601\pm 0.052} $\\
$\jdot \propto \mdot ^{2.18 \pm 0.56 }$\\
$\tau \propto (\mdot/R_\star^2)^{-1.70 \pm 0.36}$\\ 
 \hline			
\end{tabular}
\end{table} 

The stellar wind flows along open magnetic field lines, so knowing the amount of unsigned open magnetic flux $\Phi_{\rm open}$ with respect to the total observed surface flux $\Phi_0$  can be useful to predict wind properties, such as mass-loss rates and angular momentum-loss rates (cf.~Section~\ref{sec.modelcomp}). We find that the amount of open flux is approximately linearly related to the observed unsigned large-scale surface flux ($\Phi_{\rm open}\propto \Phi_0^{0.96 \pm 0.12}$, Figure~\ref{fig.fit_obs}a). 

The mass loss-rate per surface area $\mdot/R_\star^2$ correlates with the unsigned surface flux $\Phi_0$ to the power of $0.578\pm 0.087$ a relatively tight correlation (Figure~\ref{fig.fit_obs}b). Although we do not find  a tight correlation between $\jdot$ and $\Phi_0$ ($p=2.12 \pm 0.52$, Figure~\ref{fig.fit_obs}c), there is an impressive correlation between the braking time scale $\tau$ and $\Phi_0$: $\tau \propto \Phi_0^{(-1.103 \pm 0.070)}$ (Figure~\ref{fig.fit_obs}d). It is still early to argue that this correlation will hold for other spectral types and only by doing a large number of simulations we will be able to verify that. Never the less, it is interesting to note that this correlation is qualitatively expected: dynamo theories predict that fast rotators should present large surface magnetic fluxes \citep[e.g.,][]{2013SASS...39..187C} and magnetic activity is indeed observed to be strong for fast rotators \citep{2003A&A...397..147P}. In addition, theories predict that fast rotators lose angular momentum at faster rates than slow rotators \citep{1967ApJ...148..217W,1984LNP...193...49M}, therefore presenting shorter braking time scales. As the star slows down, the braking time scale becomes increasingly longer and its magnetic flux becomes smaller. Our results therefore support this picture, i.e., dM-stars with large magnetic fluxes should have shorter braking time scales. 

\subsection{Dependences with wind parameters}\label{sec.modelcomp}
\begin{figure}
\includegraphics[width=80mm]{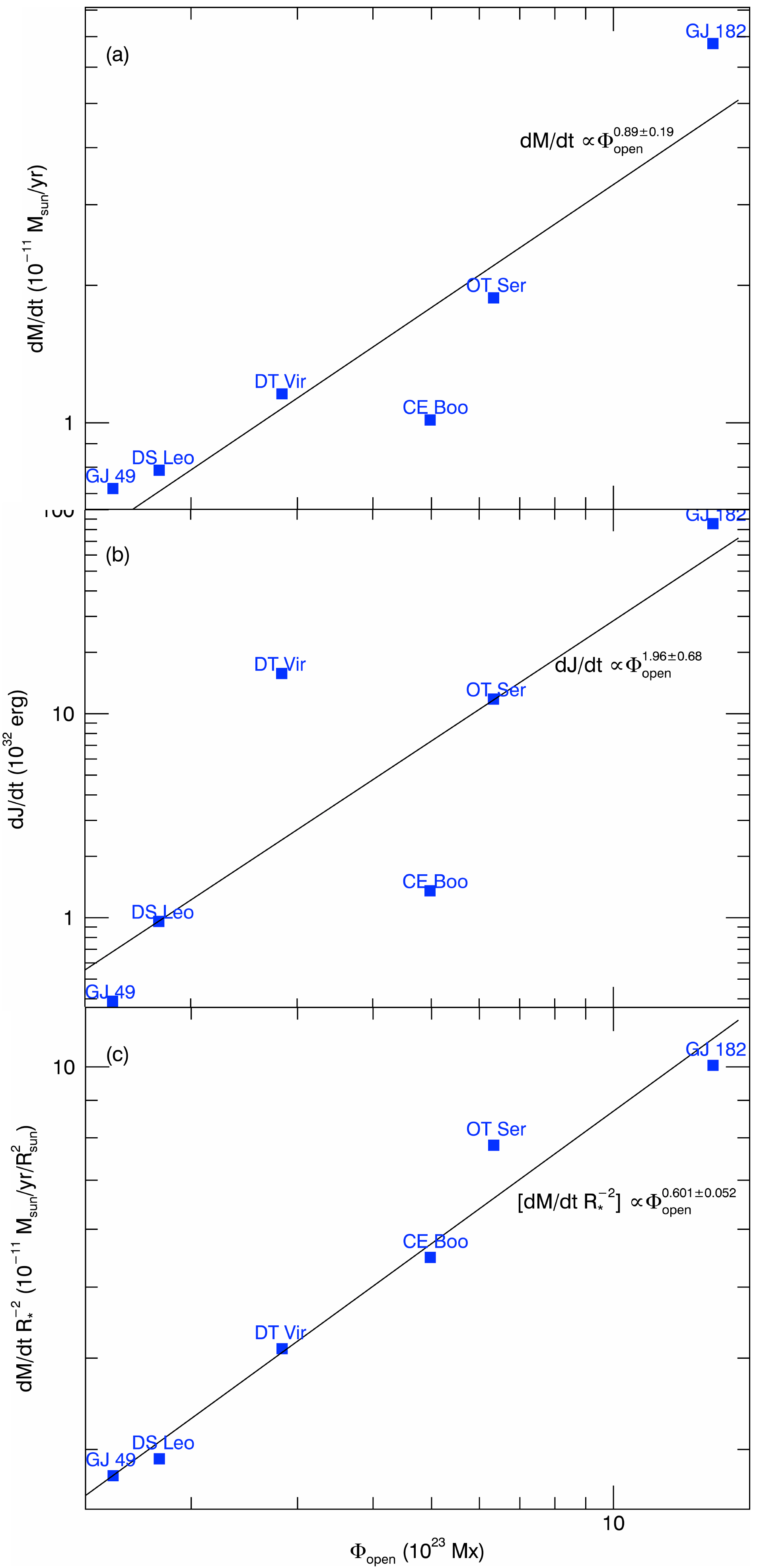}
\caption{Dependence of a few selected wind-derived parameters with the unsigned open magnetic flux $\Phi_{\rm open}$. Solid lines are power-law fits of the type $f(\Phi_{\rm opem}) \propto \Phi_{\rm opem}^{p}$, where the dependent variable $f$ is, from top to bottom, $\mdot$,  $\jdot$ and $\mdot/R_\star^2$. The power-law indexes $p$ of the fits are shown in the bottom portion of Table~{\ref{tab.fits}}.  \label{fig.fit_openflux}}
\end{figure}

\begin{figure}
\includegraphics[width=80mm]{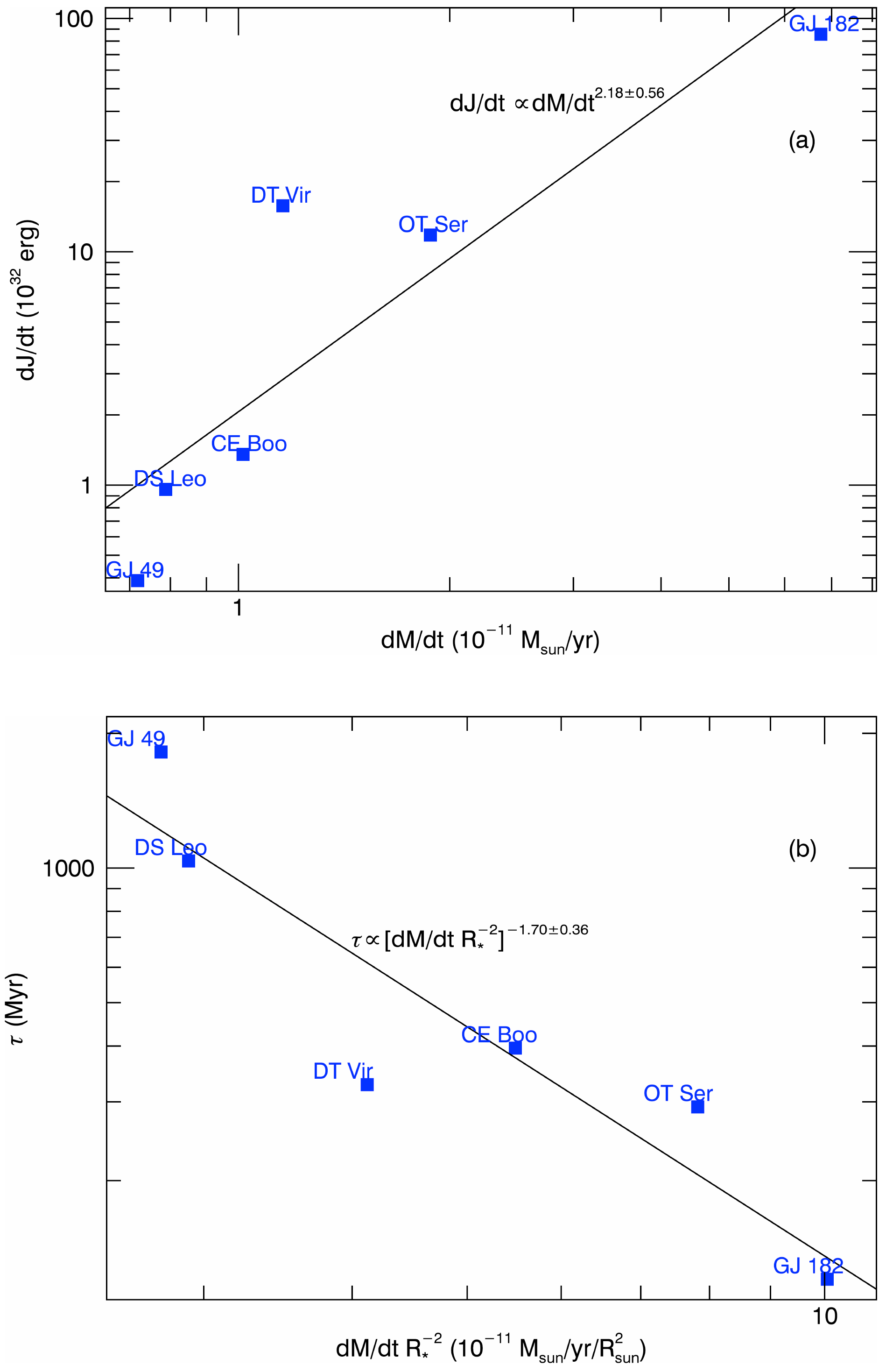}
\caption{Dependence of a few selected wind-derived parameters. Top panel shows how $\jdot$ and $\mdot$ are related and bottom panel shows relation between $\mdot/R_\star^2$ and $\tau$. Solid lines are power-law fits of the type $f(x) \propto x^{p}$. The power-law indexes $p$ of the fits are shown in the bottom portion of Table~{\ref{tab.fits}}.  \label{fig.fit_other}}
\end{figure}

Astrophysical outflows have long been studied \citep[e.g.,][]{1958ApJ...128..664P, 1967ApJ...148..217W, 1968MNRAS.138..359M, 1975ApJ...196..837N,1986ApJ...302..163L}. Several works have provided a magnetic braking formulation for computing angular momentum-loss rates of solar-type stars \citep{1993MNRAS.262..936W,2012ApJ...754L..26M,2012ApJ...746...43R},   \citet{1988ApJ...333..236K} being the currently most largely used formalism. These works generally provide how $\jdot$ depends on $\mdot$, $R_\star$, $M_\star$, $P_{\rm rot}$, and on the strength of the radial/dipolar magnetic field (usually also parameterised as a function of $P_{\rm rot}$). Due to our small set of simulations and the largely non-homogenous magnetic field topologies, it is not possible to isolate the individual dependences of the stellar parameters on $\jdot$. To numerically achieve that, a considerably large set of simulations would be required. In spite of this limitation, we have presented how a few wind-derived properties ($\mdot$, $\mdot/R_\star^2$, $\jdot$, $\Phi_{\rm open}$, $\tau$) are related to each other (lower portion of Table~\ref{tab.fits}, Figures~\ref{fig.fit_openflux} and \ref{fig.fit_other}), in a similar way as we did in Section~\ref{sec.fits_obs}. 

The increase of $\mdot$ and $\jdot$ with the unsigned open flux in slow rotators is predicted in the spherically symmetric model developed by \citet{1967ApJ...148..217W}, as $\mdot \propto \Phi_{\rm open}$ and $\jdot \propto  \Phi_{\rm open}^2$ \citep[e.g.,][]{1984LNP...193...49M}. In their model, \citet{1967ApJ...148..217W} assume a radial surface magnetic field, which develops a relatively large azimuthal component with distance, caused by  stellar wind stresses. Our model, on the other hand, incorporates realistic surface magnetic fields (derived from observations) that, similarly, are stressed by the outflowing stellar wind. From our model, we find that $\mdot \propto \Phi_{\rm open}^{(0.89\pm 0.19)}$ and $\jdot \propto \Phi_{\rm open}^{(1.96 \pm 0.68)}$ (Figures~\ref{fig.fit_openflux}a, \ref{fig.fit_openflux}b). Although the  slopes for $\mdot( \Phi_{\rm open})$ and  $\jdot( \Phi_{\rm open})$ are similar to the ones derived from the Weber-Davis model, our results show rather large scatter, which we attribute to two factors. The first cause of the large scatter is due to the wide range of field topologies (i.e., deviations from the purely radial field topology assumed in Weber-Davis model) and the second is due to dependences of $\jdot$ with multiple variables whose dependence could not be isolated (e.g., the size of the \alf\ surface; cf. Eq.~\ref{ap.eq.jdot-wd}).

A tighter correlation is found between $\mdot/R_\star^2$ and $\Phi_{\rm open}$ (Figure~\ref{fig.fit_openflux}c), with a power-law index of $p= 0.601\pm 0.052 $. In addition, our simulations suggest that $\jdot$ and $\mdot$ are correlated, although the correlation $\jdot (\mdot)$ presents a large scatter and its power-law index $p$ has a large fitting error ($p = 2.18 \pm 0.56$, cf. Fig.~\ref{fig.fit_other}a). This correlation is an important point that must be considered in works that adopt formalisms of angular momentum-loss rates. For example, \citet{1988ApJ...333..236K}  derived an analytical prescription for angular momentum losses, which, for a particular choice of magnetic field topology  (when the Alfv\'en radius is proportional to  $r^{3/2}$), $\jdot$ becomes independent of $\mdot$. This relation, although incorrect, had been widely used because it does not depend on the poorly constrained $\mdot$ for cool, low-mass stars. 

\section{Discussion}\label{sec.discussion}
\subsection{Angular momentum losses under asymmetric field geometries}\label{sec.enhance}
The presence of non-axisymmetric fields provide extra magnetic and thermal forces acting in the (asymmetric) \alf\ surface that modify the loss of angular momentum (\citealt{1970MNRAS.149..197M}, see Appendix~\ref{ap.amloss}). This is also verified in the simulations presented in \citet{2010ApJ...720.1262V}. They performed simulations of stellar winds of young Suns with different alignments between the rotation axis and the magnetic dipole axis. In those simulations, angular momentum losses were enhanced by a factor of $2$ as one goes from the aligned case to the case where the dipole is tilted by $90^{\circ}$ (i.e., as non-axisymmetry is increased).

The six early-dM stars investigated in this work have similar masses, radii, and effective temperatures, but  comprise two different groups with different rotation periods. The slowest rotating stars (GJ 49, DS Leo and CE Boo) have rotational periods of $\sim 15$~days, while the fastest rotating objects (DT Vir, OT Ser, and GJ 182) have periods of $\sim 3$~days. Despite sharing similar characteristics, members of each group present different surface magnetic field topologies and intensities. For this reason, this sample is useful for investigating the effects that different magnetic field characteristics play on stellar winds. 

We compare our results to what one would have obtained using a simplified, one-dimensional (1D) wind model. The 1D semi-analytical solution is found by assuming (i) a star with a uniformly distributed, purely radial magnetic field, whose intensity equals the average observed radial field strength ($\langle B_r\rangle=\Phi_0 (4\pi R_\star^2)^{-1}$) and (ii) a polytropic wind, with the same base density, temperature and $\gamma$ as those adopted in our 3D simulations  (i.e., the hydrodynamical quantities are as in the initial state of the simulation). Note that the 1D wind solution we construct is exact for non-rotating stars and we expect some deviations for slowly rotating stars (such as the ones in our sample). From assumptions (i) and (ii), we are then able to calculate the spherical radius ($r_{A, {\rm 1D}}$) of the Alfv\'en surface of this simplified wind model and also its mass-loss rate ($\mdot_{\rm 1D}$). We calculate the angular momentum-loss rates as in \citet{1967ApJ...148..217W}
\begin{equation}
\jdot_{\rm 1D} = \frac23 \Omega_\star r_{A, {\rm 1D}}^2 \mdot_{\rm 1D} . 
\end{equation}
Table~\ref{table2} shows how the angular momentum-loss rate $\jdot_{\rm 1D}$ predicted by a simplified 1D model compares to the value of $\jdot$ derived from our 3D simulations. We find that $\jdot/\jdot_{\rm 1D} $ ranges from 0.15 to 0.5. One reason why the simplified 1D model over-predicts angular momentum loss rates is because assumption (i) implies that all the surface magnetic field contributes to the wind in the 1D simplified model, while in the 3D solution, only the open field lines participate in the angular momentum removal. 

To isolate the effects that different field topologies might have on $\jdot$, it is interesting to compare stars with similar rotation rates and observed surface magnetic fluxes $\Phi_0$. Comparison between GJ~49 and DS~Leo shows that DS~Leo has $\jdot$ that is larger by factor of $2.5$, and comparison between OT~Ser and DT~Vir, shows that DT~Vir presents $\jdot$ that is a factor $1.3$ larger. Our results demonstrate that different field topologies indeed affects the amount of angular momentum lost in the wind, as different magnetic field intensities and topologies contribute differently to extraction of angular momentum.

\subsection{Effects on rotational evolution}\label{sec.rotev} 
Although a factor of a few in $\jdot$ seems to be a small difference, if 3D effects (due to field topology) are not properly accounted in rotational evolution models, this small deviation can lead to an error in predicting the rotation rates of dM stars. For example, if the assumed angular momentum-loss rates are smaller than the `real' ones, rotational evolution models end up predicting increasingly larger rotation rates with time and therefore an excess of fast rotators. It is difficult to quantify the error in the predicted rotation rates, since $\jdot$ is a complex function of many variables, but a rough estimate is presented next. If we take an angular momentum loss rate that depends on the angular velocity to some power $q$ ($\jdot =-c \Omega_\star^q$, for a given coefficient $c$), one can show that, for large $t$, $\Omega_\star \to [c(q-1)t]^{1/(-q+1)}$, where we assumed that $J\propto M_\star R_\star^2 \Omega_\star$ and that $M_\star$ and $R_\star$ are roughly constants (e.g., low-mass stars in the main-sequence phase). Therefore, a factor of $4$ excess in the coefficient $c$ predicts rotation rates that are different by a factor of $4^{1/(-q+1)}$. Note that for $q=3$, the Skumanich's rotation-age relation ($\Omega_\star \propto t^{-1/2}$) is recovered and, in that case, overestimating the angular momentum loss rates by a factor of $4$ predicts rotation rates that are small by a factor $2$. 

Recently, \citet{2013MNRAS.432.1203M} found that the slope of the upper envelope of the period--mass distribution (which is defined by the slowest rotating stars) changes sign at masses around $0.55 ~M_\odot$ and that for $M_\star \lesssim 0.55 ~M_\odot$ the period of the slowest rotators rises with decreasing mass. Simplified analytical models derived by \citet{2012ApJ...746...43R} are able to explain the slow rotation rates of dM stars by assuming that the rotation rate at which activity saturation occurs is much larger than the rotation rate at which higher mass stars saturate (cf.~Section~\ref{sec.intro}). An alternative, or perhaps an additional, explanation for the rise in period of the upper envelope of the period--mass relation found by \citet{2013MNRAS.432.1203M} could also be explained by different magnetic field topologies. 

\subsection{Effects on planets}
\subsubsection{Galatic cosmic rays}
Cosmic rays play important effects on the chemistry and ionisation of planetary atmospheres \citep{2013P&SS...77..152H,2013arXiv1307.3257R}. They could also be a source of genetic mutations in organisms \citep{2012arXiv1211.3962A}. Therefore, the impact of cosmic ray flux on exoplanets may have important implications for both atmospheric characterisation and habitability.

The flux of cosmic rays that impact the Earth is modulated over the solar cycle. \citet{2006ApJ...644..638W} found that the non-axisymmetric component of the total open solar magnetic flux is inversely correlated to the cosmic-ray rate. By analogy, if the non-axisymmetric component of the stellar magnetic field is able to reduce the flux of cosmic rays reaching the planet, then we would expect that planets orbiting stars with largely non-axisymmetric fields would be more shielded from galactic cosmic rays, independently of the planet's own shielding mechanism (such as the ones provides by a thick atmosphere or large magnetosphere). 

From reconstructions of stellar magnetic fields using the ZDI technique, it is possible to separate the axisymmetric part of the surface field from the non-axisymmetric one \citep{2006MNRAS.370..629D}. However, it is not obvious if the non-axisymmetric field should maintain its surface characteristics at large distances. Here we calculate the unsigned magnetic fluxes considering both the axi- and non-axisymmetric components of the magnetic field from the results of our simulations, following the approach of \citet{2006ApJ...644..638W}. Their approach considers a potential field extrapolation, where the axisymmetric component of the magnetic field is obtained by averaging the contribution of the spherical harmonics of order $m=0$ over longitude $\varphi$. Although the magnetic field derived in our simulations is not potential, we adopt a similar approach and define the axi-symmetric component of the magnetic field as 
\begin{equation}
B_r^{\rm axi} (r,\theta) = \frac{1}{2\pi} \oint B_r (r,\theta, \varphi)  d\varphi . 
\end{equation}
The corresponding unsigned axisymmetric magnetic flux is 
\begin{equation}
\Phi_{\rm axi}=\oint  |B_r^{\rm axi} | dS_{\rm sph} = \oint r^2 \left( \left| \oint B_r d\varphi \right| \right) \sin \theta d \theta .
\end{equation}

In our computations, we take a sphere close to the outer edge of our simulations ($r\sim 20~R_\star$). We find that DT~Vir, DS~Leo and GJ~182 have the smallest ratio of $\Phi_{\rm axi}/\Phi_{\rm open}$ (0.18, 0.52, and 0.65 respectively). On the other hand, CE~Boo, GJ~49 and OT~Ser are very axisymmetric and present the largest values of $\Phi_{\rm axi}/\Phi_{\rm open}$ (0.95, 0.88 and 0.97, respectively). Although these flux ratios are not identical to the axisymmetric-to-total magnetic energy ratios observationally derived \citep[][cf. $f_{\rm axi}$  in Table~\ref{table}]{2008MNRAS.390..545D}, the trend is similar to the observed one. This means that the axisymmetric magnetic field topologies that are observed can be used to characterise the degree of axisymmetric field at large distances from the star. Therefore, if cosmic ray shielding is more efficient in planets orbiting stars whose magnetic fields are more non-axisymmetric, then planets orbiting DT~Vir, DS~Leo and GJ~182 should be the most effectively shielded planets from galactic cosmic rays. Detailed computations of the propagation of cosmic rays such as those performed in \citet{2006AN....327..866S, 2012ApJ...760...85C,2013ApJ...772....5C} should provide better constraints of the effectiveness of the shielding of cosmic rays due to the non-axisymmetry of the host-star magnetic fields.

\subsubsection{Planetary magnetospheres}
In Section~\ref{sec.massflux}, we showed that the mass flux profile is essentially modulated by the local value of $|B_r|$ and that, at least within our simulation domain, we find that the more non-axisymmetric topology of the stellar magnetic field results in more asymmetric mass fluxes distribution.

\begin{figure*}
\includegraphics[width=56mm]{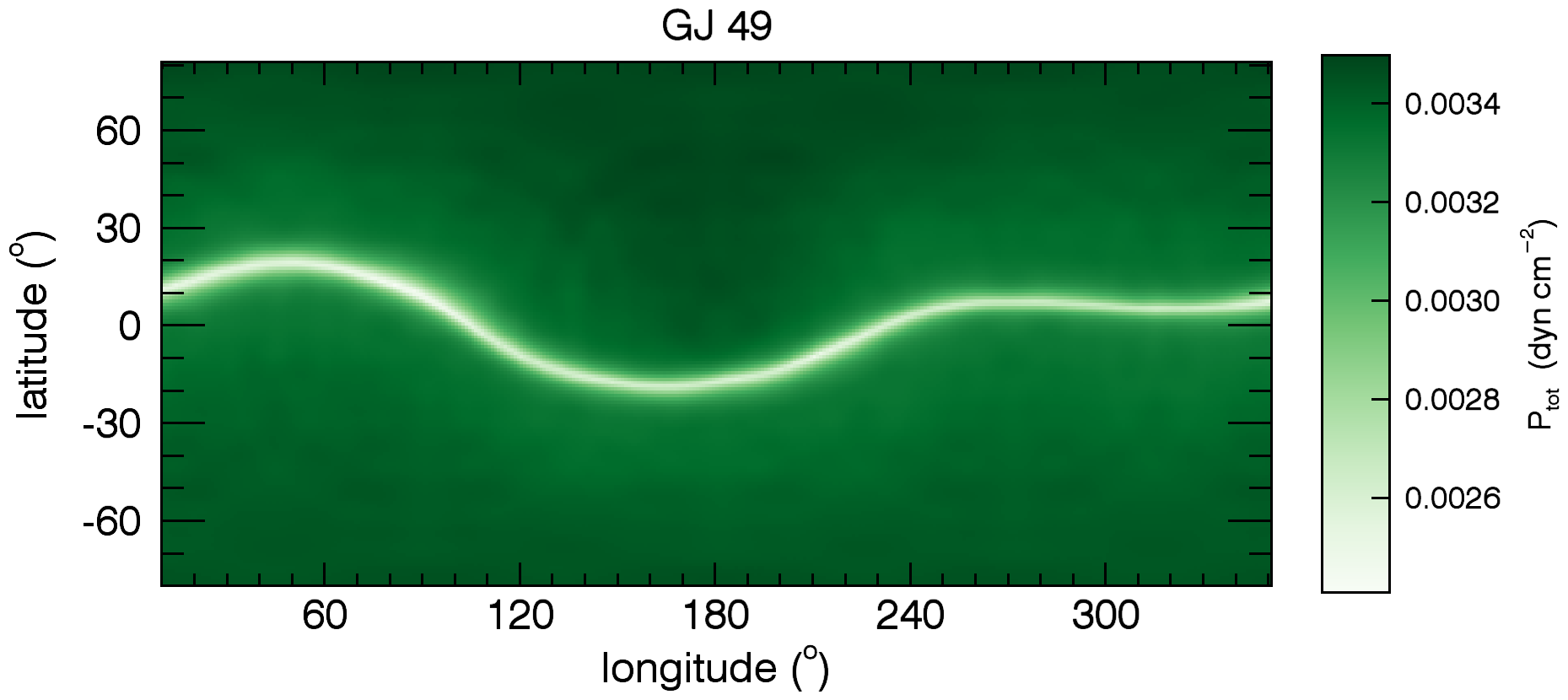}
\includegraphics[width=56mm]{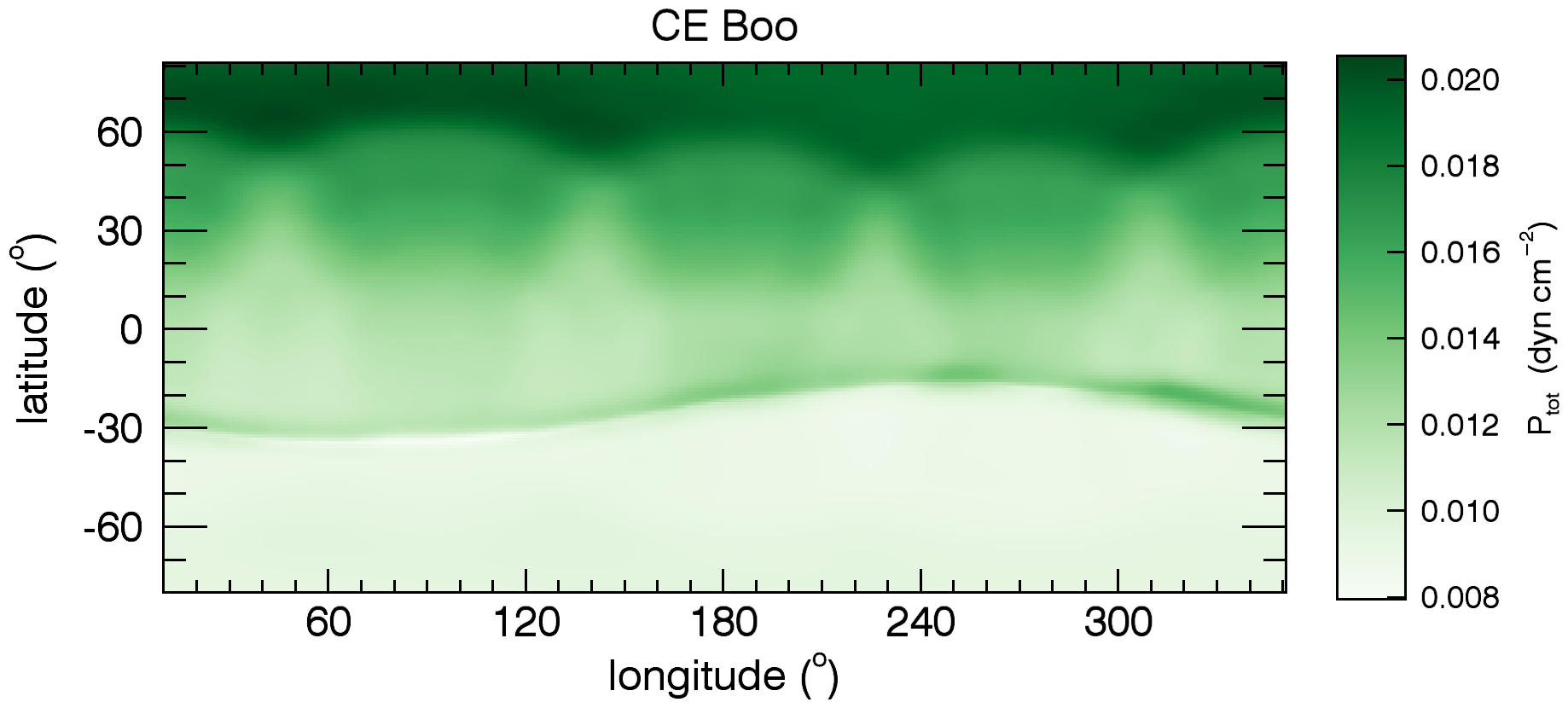}
\includegraphics[width=56mm]{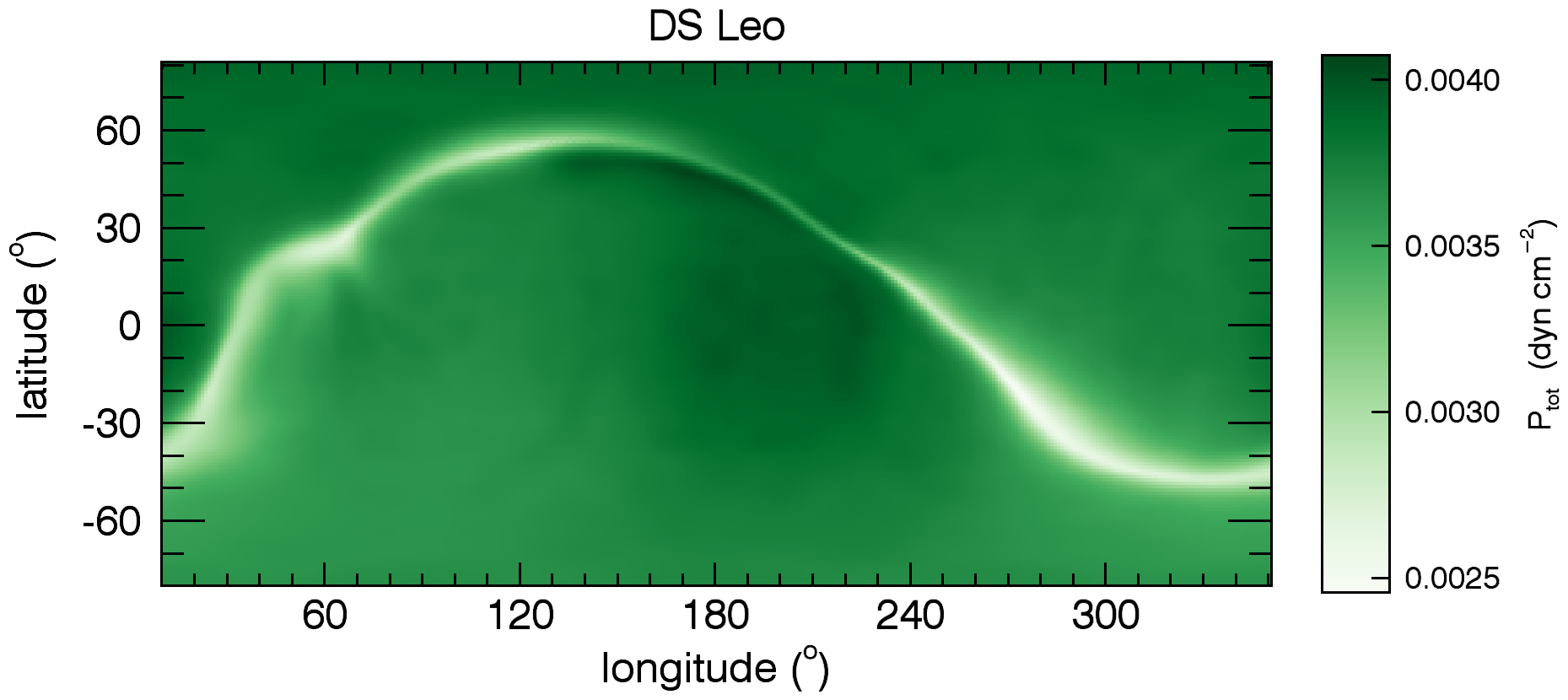}\\
\includegraphics[width=56mm]{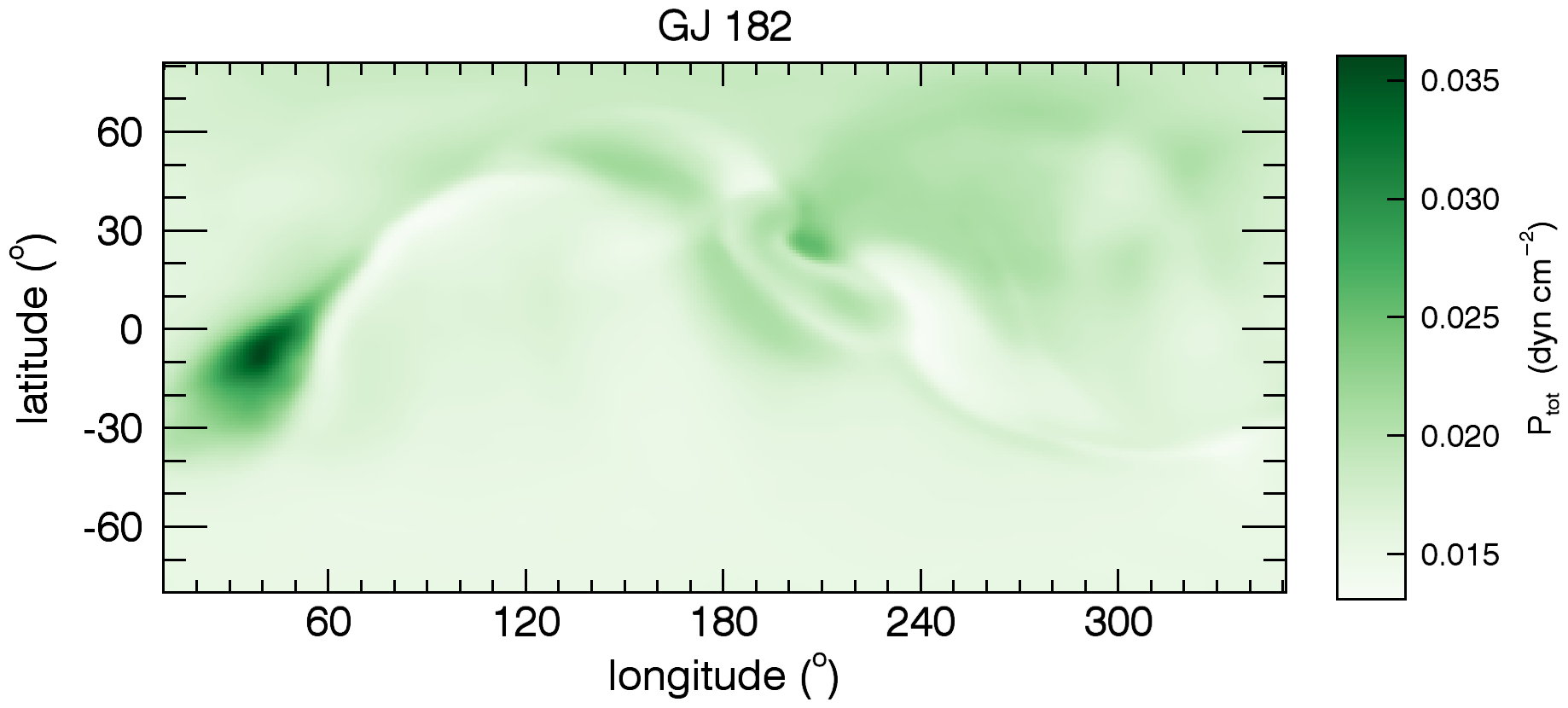}
\includegraphics[width=56mm]{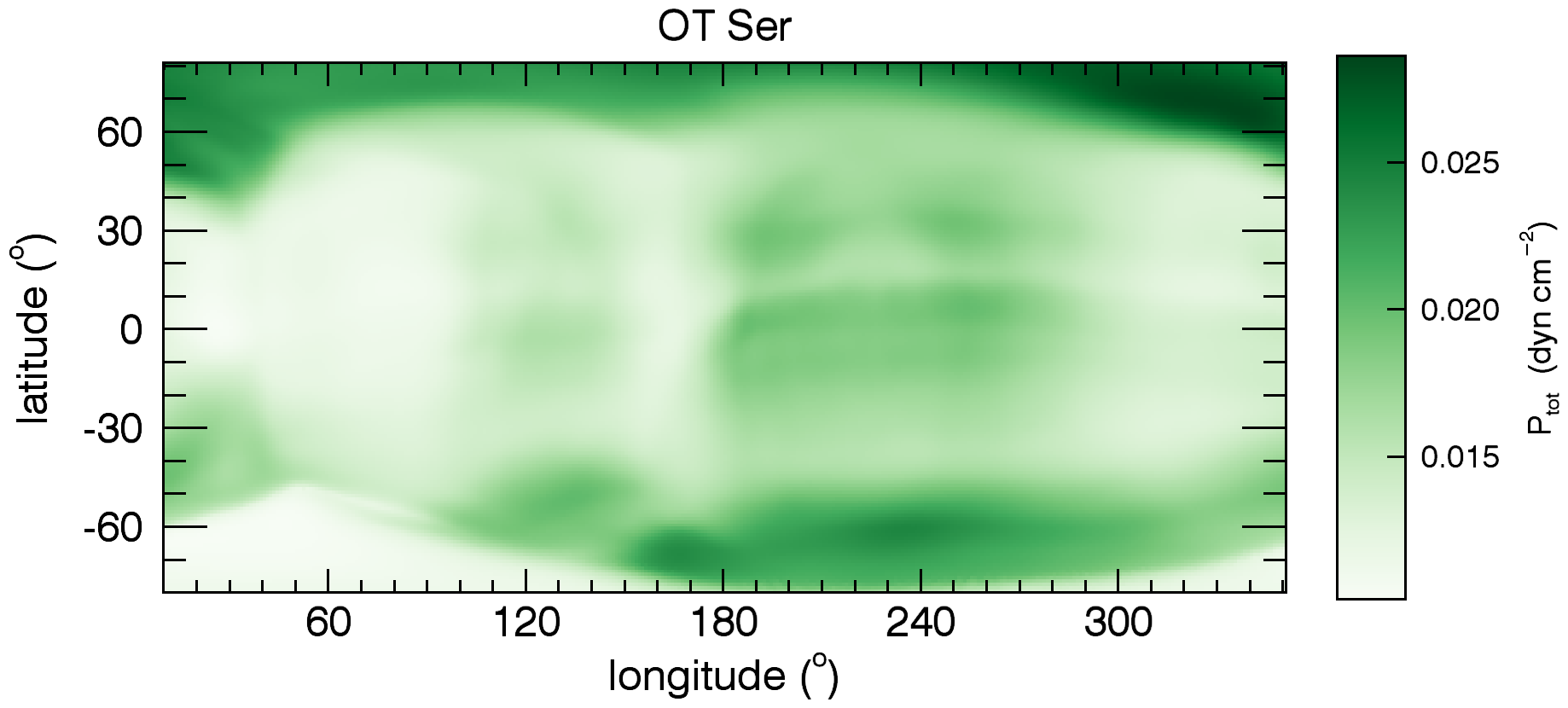}
\includegraphics[width=56mm]{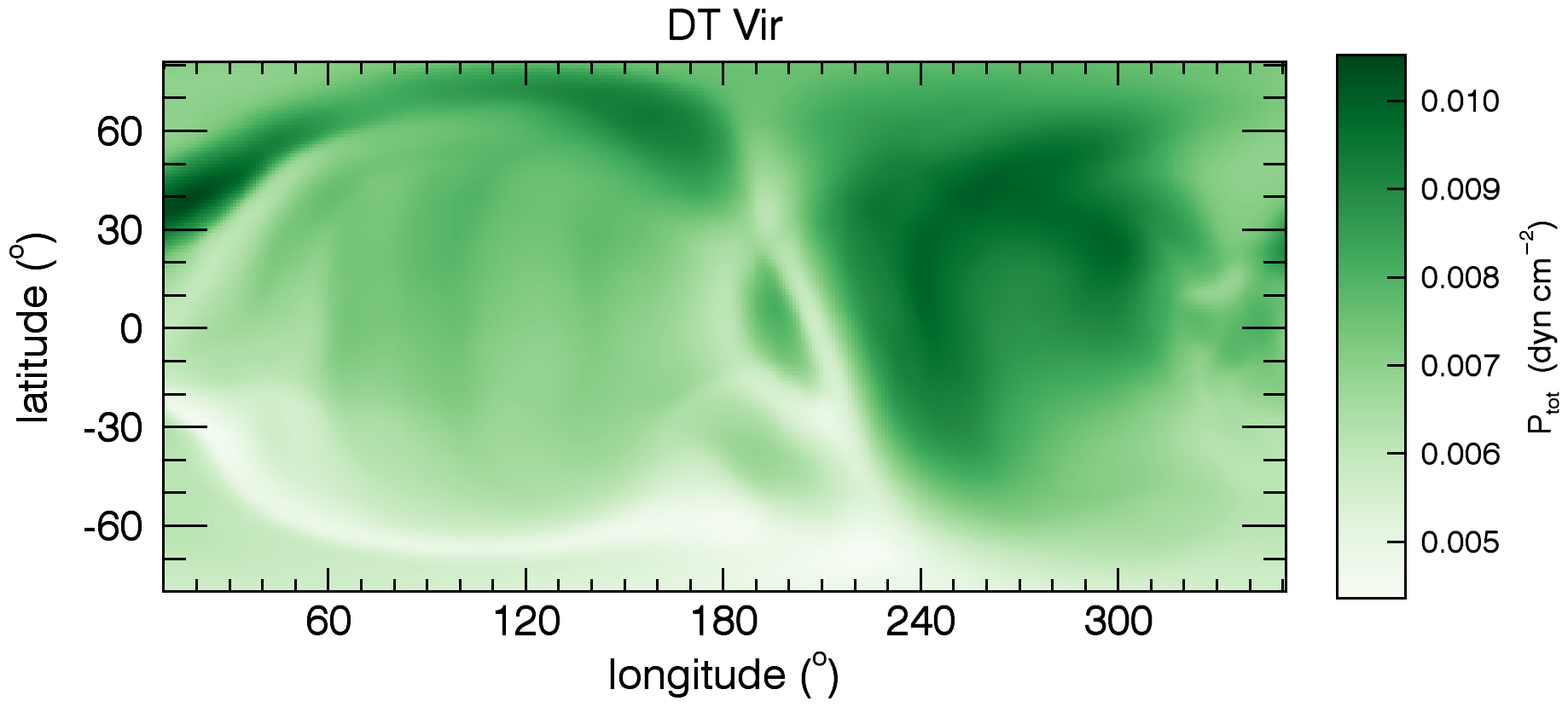}\\
\caption{Distribution of the stellar wind total pressure $p_{\rm tot}$ at a spherical surface of radius $\sim 19~R_\star$ (close to the edge of our simulation domain). The more asymmetric topology of the stellar magnetic field results in more asymmetric $p_{\rm tot}$. Because planetary magnetospheric sizes $r_M$ are set by pressure equilibrium between the planet's magnetic field and $p_{\rm tot}$, as the planet interacts with the wind of its host star along its orbital path, $r_M$ becomes smaller (larger) when the external $p_{\rm tot}$ is larger (smaller). 
\label{fig.ptot}}
\end{figure*}

Likewise, we found that the stellar wind total pressure $p_{\rm tot}$ (i.e., the sum of thermal, magnetic and ram pressures)  is also modulated by $|B_r|$ and similarly, the more non-axisymmetric topology of the stellar magnetic field produces more asymmetric distributions of $p_{\rm tot}$. Figure~\ref{fig.ptot} shows the distribution of $p_{\rm tot}$ for a sphere located at the outer edge of our simulation domain (at $r \sim 19~R_\star$), where we can see that variations of up to a factor of $3$ is obtained. 

Considering a magnetised planet in orbit around a star, pressure balance between the wind total pressure and the planet total pressure requires that, at a the magnetopause distance $r_M$ from the planet,
\begin{equation}\label{eq.equilibrium}
 p_{\rm tot} = \frac{B_{{p},r_M}^2}{8\pi} + p_{p} ,
\end{equation}
where $B_{{p},r_M}$ is the planetary magnetic field intensity at a distance $r_M$ from the planet centre and $p_p$ is its thermal pressure. Along their orbital paths, planets interact with the wind of their host stars. By probing regions with different $p_{\rm tot}$, the magnetospheric sizes of planets react accordingly, becoming smaller (larger) when the external $p_{\rm tot}$ is larger (smaller).   \citet{2011MNRAS.414.1573V} investigated other effects that could cause variability in planetary magnetospheric sizes. 

Neglecting the thermal pressure of the planet and assuming the planetary magnetic field is dipolar, we have that $B_{{p},r_M} = B_{p, {\rm eq}} (R_p/r_M)^3$, where $R_p$ is the planetary radius and $B_{p, {\rm eq}}$ its surface magnetic field at the equator. For a planetary dipolar axis aligned with the rotation axis of the star, the magnetospheric size of the planet is given by
\begin{equation}\label{eq.r_M}
 \frac{r_M}{R_p} = \left[ \frac{B_{p, {\rm eq}}^2}{8 \pi p_{\rm tot}} \right]^{1/6}.
\end{equation}
For example, a factor of $3$ difference in $p_{\rm tot}$ along its orbit will cause the planet magnetosphere to reduce/expand by $20$~percent. 

The angle between the shock normal and the tangent of a circular orbit is defined as \citep{2010ApJ...722L.168V}
\begin{equation}
\theta_{\rm shock} = \arctan \left( \frac{u_r}{|u_K-u_\varphi|}\right),
\end{equation}
where $u_K$ is the Keplerian velocity of the planet. Along its orbital path, the planet probe regions of the wind with different velocities ($u_r$ and $u_\varphi$ in the equation above). Therefore, in addition to magnetospheric size variations, the orientation of the bow shock that forms surrounding planetary magnetospheres  also change along the planetary orbit, as a consequence of asymmetric stellar magnetic field distributions (see also \citealt{2013arXiv1309.2938L}, for the specific case of HD~189733b).

\section{Summary and CONCLUSIONS}\label{sec.conclusions}
In this work, we investigated how the stellar winds of early-dM stars respond to variations in their surface magnetic field characteristics.  We presented MHD numerical simulations of the wind of six M dwarf (dM) stars with spectral types M0 to M2.5, for which observationally derived surface magnetic field maps exist \citep{2008MNRAS.390..545D}. To account for the observed three-dimensional (3D) nature of their magnetic fields, 3D stellar wind models are required. Starting from an initial potential magnetic field configuration and a thermally-driven wind, the system is evolved in time, resulting in a self-consistent interaction of the wind particles and the magnetic field lines. We note that all the simulations were performed with the same grid resolution, boundary and initial conditions. They also adopted the same coronal base density and temperature. Provided that these quantities are similar for other stars, our results should be extendable to other spectral types (e.g., to mid- and late-dM stars). Masses, radii, rotation periods and surface magnetic field distributions were adopted as shown in Table~\ref{table} and Figure~\ref{fig.maps}, following the results published \citet{2008MNRAS.390..545D}. The summary of our simulation results are found in Table~\ref{table2}. 

Contrary to the results obtained on wind models with axisymmetric magnetic fields, the \alf\ surfaces of the objects investigated here have irregular, asymmetric shapes, which can only be captured by fully 3D models. We found that the more non-axisymmetric topology of the stellar magnetic field results in more asymmetric mass fluxes. We also found that there is no preferred colatitude that contributes more to mass loss, as the mass flux is maximum at different colatitudes for different stars. We note that latitudinal and longitudinal variations in mass flux should also affect the distances and shapes of astropauses, which would lack symmetry due to the asymmetric nature of the stellar magnetic field. 

We also computed the rate $\jdot$ of angular momentum carried by the stellar wind and found that it varies by more than two orders of magnitude among our targets (Fig.~\ref{fig.meridional}b). The variations we found in $\dot{J}$ are related not only to differences in rotation periods, but also to changes in the topology and intensity of the magnetic fields. In spite of the diversity in the magnetic field topology, we found that the stellar wind flow at equatorial regions carries most of the stellar angular momentum for the stars studied in this work. Our simulations suggested that the complexity of the magnetic field can play an important role in the angular momentum evolution of the star, as different magnetic field intensities and topologies contribute differently to extraction of stellar angular  momentum. Different magnetic field topologies are therefore a plausible explanation for a large distribution of periods in field dM stars, as has been found recently \citep{2011ApJ...727...56I,2013MNRAS.432.1203M}. 

The lack of symmetry in the topology of the stellar field can also affect any orbiting planet. The flux of cosmic rays that impact the Earth is modulated over the solar cycle. \citet{2006ApJ...644..638W} found that the non-axisymmetric component of the total open solar magnetic flux is inversely correlated to the cosmic-ray rate. Therefore, if cosmic ray shielding is more efficient in planets orbiting stars whose magnetic fields are more non-axisymmetric, then planets orbiting stars like DT~Vir, DS~Leo and GJ~182, which have largely non-axisymmetric fields, should be the most shielded planets from galactic cosmic rays, even if the planets lack  protective thick atmosphere or large magnetosphere of their own. 

The size of the magnetosphere $r_M$ of a planet (Eq.~(\ref{eq.r_M})) is set by pressure equilibrium between the planet's magnetic field and the stellar wind total pressure $p_{\rm tot}$ (i.e., the sum of thermal, magnetic and ram pressures). Similarly to the mass-flux, we found that $p_{\rm tot}$ is essentially modulated by the local value of $|B_r|$, which presents similar structures as the observed surface $|B_r|$. Therefore, as the planet interact with the wind of its host star along its orbital path, it probes regions with different $p_{\rm tot}$. As a consequence, the magnetospheric sizes of planets react accordingly, becoming smaller (larger) when the external $p_{\rm tot}$ is larger (smaller). For example, a factor of $3$ difference in $p_{\rm tot}$ (typical of what was found in our simulations) along its orbit will cause the planet magnetosphere to reduce/expand by $20$~percent. In addition to magnetospheric size variations, the orientation of the bow shock that forms surrounding planetary magnetospheres also change along the planetary orbit, as a consequence of asymmetric stellar magnetic field distributions.

\section*{Acknowledgements}
AAV acknowledges support from a Royal Astronomical Society Fellowship. JM acknowledges support from a fellowship of the Alexander von Humboldt foundation. This work used the DiRAC Data Analytic system at the University of Cambridge, operated by the University of Cambridge High Performance Computing Service on behalf of the STFC DiRAC HPC Facility (www.dirac.ac.uk). This equipment was funded by BIS National E-infrastructure capital grant (ST/K001590/1), STFC capital grants ST/H008861/1 and ST/H00887X/1, and STFC DiRAC Operations grant ST/K00333X/1. DiRAC is part of the National E-Infrastructure.

\def\aj{{AJ}}                   
\def\araa{{ARA\&A}}             
\def\apj{{ApJ}}                 
\def\apjl{{ApJ}}                
\def\apjs{{ApJS}}               
\def\ao{{Appl.~Opt.}}           
\def\apss{{Ap\&SS}}             
\def\aap{{A\&A}}                
\def\aapr{{A\&A~Rev.}}          
\def\aaps{{A\&AS}}              
\def\azh{{AZh}}                 
\def\baas{{BAAS}}               
\def\jrasc{{JRASC}}             
\def\memras{{MmRAS}}            
\def\mnras{{MNRAS}}             
\def\pra{{Phys.~Rev.~A}}        
\def\prb{{Phys.~Rev.~B}}        
\def\prc{{Phys.~Rev.~C}}        
\def\prd{{Phys.~Rev.~D}}        
\def\pre{{Phys.~Rev.~E}}        
\def\prl{{Phys.~Rev.~Lett.}}    
\def\pasp{{PASP}}               
\def\pasj{{PASJ}}               
\def\qjras{{QJRAS}}             
\def\skytel{{S\&T}}             
\def\solphys{{Sol.~Phys.}}      
\def\sovast{{Soviet~Ast.}}      
\def\ssr{{Space~Sci.~Rev.}}     
\def\zap{{ZAp}}                 
\def\nat{{Nature}}              
\def\iaucirc{{IAU~Circ.}}       
\def\aplett{{Astrophys.~Lett.}} 
\def\apspr{{Astrophys.~Space~Phys.~Res.}}   
\def\bain{{Bull.~Astron.~Inst.~Netherlands}}    
\def\fcp{{Fund.~Cosmic~Phys.}}  
\def\gca{{Geochim.~Cosmochim.~Acta}}        
\def\grl{{Geophys.~Res.~Lett.}} 
\def\jcp{{J.~Chem.~Phys.}}      
\def\jgr{{J.~Geophys.~Res.}}    
\def\jqsrt{{J.~Quant.~Spec.~Radiat.~Transf.}}   
\def\memsai{{Mem.~Soc.~Astron.~Italiana}}   
\def\nphysa{{Nucl.~Phys.~A}}    
\def\physrep{{Phys.~Rep.}}      
\def\physscr{{Phys.~Scr}}       
\def\planss{{Planet.~Space~Sci.}}           
\def\procspie{{Proc.~SPIE}}     

\let\astap=\aap
\let\apjlett=\apjl
\let\apjsupp=\apjs
\let\applopt=\ao
\let\mnrasl=\mnras

\appendix
\section{Angular momentum losses in stars with non-axisymmetric field topologies}\label{ap.amloss}
To evaluate the angular momentum-loss rate carried by the winds simulated in this work, we compute the torque $\dot{\bf J}$ applied on the star by the outflow of magnetised winds. In this Appendix, we present a step-by-step derivation of the angular momentum-loss rate considering a system that lacks symmetry. The derivation performed next follows very closely the one presented in \citet{1970MNRAS.149..197M} and \citet{1999stma.book.....M}. 

The $i$-component of the torque density is given by $l_i = ({\bf r} \times {\bf f})_i$, where ${\bf f}$ is the force per unit volume
\begin{equation}
{\bf f} = - {\bf \nabla \cdot T} 
\end{equation}
where in tensor form $T_{kl}$ is given by \citep{1999stma.book.....M}
\begin{equation}
T_{kl} = \left( \frac{B^2}{8\pi} + P \right)\delta_{kl} - \frac{B_k B_l}{4 \pi}  + \rho V_l (V_k + (\boldsymbol{\Omega_\star} \times {\bf r})_k) \, ,
\end{equation}
where ${\bf V} = {\bf u} - \boldsymbol{\Omega_\star}\times {\bf r}$ is the velocity vector in the frame rotating with angular velocity $\boldsymbol{\Omega_\star}$ and $\bf u$ is the velocity in the inertial reference frame. The outflow per unit area of the $i$-component of the angular momentum across a volume $V$ bounded by a closed surface $S$ is
\begin{eqnarray}
-\dot{J}_i=-\int_V{l_i dV} &=& \int_V \epsilon_{ijk} x_j \frac{d T_{kl}}{d x_l} dV \nonumber \\ 
&=& \int_V \frac{d}{d x_l}(\epsilon_{ijk} x_j T_{kl}) dV \nonumber \\
&=& \oint_S \epsilon_{ijk} x_j T_{kl} n_l dS
\end{eqnarray}
where $n_i$ is the normal vector to the surface $S$, $\epsilon_{ijk}$ the Levi-Civita permutation symbol and $x_i$ is the coordinate vector. We used the property that $T_{kl}$ is symmetric from the first to the second line and the divergence theorem from the second to the third line. The subscripts `1', `2' and `3' denote, respectively, the $x$, $y$ and $z$ components of a given vector/tensor. We focus only on the $z$-component of $\dot{\bf J}$, as it is the one responsible for the star's rotational braking (as $\boldsymbol{\Omega_\star}$ points in the $z$-direction). Therefore, the $z$-component of the angular momentum carried by the wind is 
\begin{equation}
\dot{J}_3 = \oint_S \epsilon_{3jk} x_j T_{kl} n_l dS= \oint_S (- x_2 T_{1l} +  x_1 T_{2l})n_ldS ,
\end{equation}
where we dropped the minus sign ahead of $\jdot_3$, but remind the reader that it refers to the angular momentum that is {\it lost}. After rearranging terms, we have
\begin{eqnarray}\label{ap.eq.jdot}
\dot{J}_3 &=& \oint_S^{T_1} (-x_1 B_2 + x_2 B_1) \left( \frac{\bf B \cdot n}{4 \pi} \right) {\rm d} S  \nonumber \\ 
&+& \oint_S^{T_2}  (x_1 n_2 - x_2 n_1) \left( P+ \frac{B^2}{8 \pi} \right) {\rm d} S  \nonumber  \\ 
&+& \oint_S^{T_3} \Omega_\star \varpi^2 \rho ({\bf V \cdot n}) {\rm d} S  \nonumber  \\ 
&+& \oint_S^{T_4}  (x_1 V_2 - x_2 V_1) \rho ({\bf V \cdot n})  {\rm d} S ,
\end{eqnarray}
where $\varpi=(x_1^2+x_2^2)^{1/2}$ is the cylindrical radius and $T_1$ to $T_4$ denote each of the four terms of this equation, which will be discussed below. Eq.~(\ref{ap.eq.jdot}) is valid for any closed surface that encloses the star. In particular, because ${\bf B \parallel V}$, at the Alfv\'en surface $S_A$, $B/V=\sqrt{4 \pi \rho}$ and  it can be shown that $T_1=-T_4$. Thus, Equation~(\ref{ap.eq.jdot}) simplifies to \citep{1999stma.book.....M}
\begin{eqnarray}\label{ap.eq.jdot_sa}
\dot{J}_3 &=& \oint_{S_A} \left[ ({\bf r_A \times n_A})_3 \left(P_A + \frac{B_A^2}{8\pi}\right) \right. \nonumber  \\  &+& \left. \rho_A ({\bf V_A \cdot n_A})  \Omega_\star \varpi_A^2 \right] dS_A,
\end{eqnarray} 
where the index `$A$' is used to remind us that the variable is computed at the \alf\ surface. The second term in Eq.~(\ref{ap.eq.jdot_sa}) is the effective corotation term, which is the only non-null term under spherical symmetry. The first term is the moment about the centre of the star of the thermal and magnetic pressures acting on the (asymmetric) \alf\ surface. Note that the presence of non-axisymmetry provides extra forces acting on the \alf\ surfaces that modify the loss of angular momentum. 
It is straightforward to show that under {\it spherical symmetry}, Eq.~(\ref{ap.eq.jdot_sa}) reduces to the known relation derived  by \citet{1967ApJ...148..217W} 
\begin{eqnarray}\label{ap.eq.jdot-wd}
\dot{J}_{\rm WD} = \oint_{S_A} \rho ({\bf V \cdot n})  \Omega_\star \varpi_A^2 {\rm d} { S_A}  = \frac23 \dot{M} r_A^2 \Omega_\star,
\end{eqnarray} 
where $r_A$ is the radius of the spherical Alfv\'en surface. We stress here that previous equation is only valid for systems with axial symmetry and Eq.~(\ref{ap.eq.jdot_sa}) or (\ref{ap.eq.jdot_3}) below should be used in asymmetric field configurations. 

Because our integration is done numerically and the \alf\ surface in our simulations are usually quite irregular (due to the asymmetric nature of the magnetic field distribution, cf.~Fig.~\ref{fig.IC-SS}), to reduce numerical errors in our computation, we integrate $\dot{J}_3$ over spherical surfaces at different distances from the star. In this case, the term $T_2$ in Eq.~(\ref{ap.eq.jdot}) is null ($x_1 n_2 - x_2 n_1\equiv ({\bf r \times n})_3=0$) and we are left with a contribution from the magnetic torque (term $T_1$) and a contribution from the angular momentum of the material (terms $T_3+T_4$). Rearranging terms, we have
\begin{eqnarray}\label{ap.eq.jdot_3}
\dot{J}_3 &=& \oint_{S_{\rm sph}} \left[ (-x_1 B_2 + x_2 B_1) \left( \frac{\bf B \cdot n}{4 \pi} \right) \right .\nonumber \\
&+& \left. (\Omega_\star \varpi^2 + x V_2 - x_2 V_1) \rho ({\bf V \cdot n}) \right]  {\rm d} S _{\rm sph} \nonumber \\
&=& \oint_{S_{\rm sph}} \left[  - \frac{\varpi B_\varphi B_r}{4 \pi} + \varpi u_\varphi \rho u_r \right]  {\rm d} S _{\rm sph} ,
\end{eqnarray} 
where in the last equality we used spherical coordinates at the inertial reference frame. 

Equations (\ref{ap.eq.jdot_sa}) and (\ref{ap.eq.jdot_3}) are mathematically equivalent. Because the former one requires the computation of the normal vectors to the highly irregular Alfv\'en surfaces, Equation~(\ref{ap.eq.jdot_3}) is computationally more efficient in systems that lack symmetry.

\bsp

\label{lastpage}
\end{document}